\documentclass{ws-ijmpa}
\usepackage[super,compress]{cite}
\usepackage{graphicx}
\usepackage{hyperref}
\usepackage{dsfont}

\newcommand{\be}{\begin{equation}}
\newcommand{\en}{\end{equation}}
\newcommand{\bii}{\begin{itemize}}
\newcommand{\eii}{\end{itemize}}
\newcommand{\eeq}{\end{equation}}
\newcommand{\bea}{\begin{eqnarray}}
\newcommand{\ena}{\end{eqnarray}}
\newcommand{\dlangle}{\left\langle \kern-.17em \left\langle}
\newcommand{\drangle}{\right\rangle \kern-.17em \right\rangle}
\newcommand{\hbo}{\hbox to 1 true cm {\hfill } }

\newcommand{\Tr}{\hbox{Tr}}
\newcommand{\e}{\mathrm{e}}
\newcommand{\lb}{\langle \kern-.17em \langle} 
\newcommand{\rb}{\rangle \kern-.17em \rangle }
\newcommand{\dd}{{\rm d}}
\newcommand{\DE}{\delta E}

\begin{document}
\markboth{C Gattringer, K Langfeld}{The Sign Problem}

%
\catchline{}{}{}{}{}
%

\title{Approaches to the sign problem in lattice field theory}

\author{Christof Gattringer}

\address{Institut f\"ur Physik, Universit\"at Graz,
8010 Graz, \"Osterreich, \\
christof.gattringer@uni-graz.at}

\author{Kurt Langfeld}

\address{Centre for Mathematical Sciences, Plymouth University,
  Plymouth, PL4 8AA, UK, \\  
kurt.langfeld@plymouth.ac.uk}

\maketitle

\begin{history}
\received{Day Month Year}
\revised{Day Month Year}
\end{history}

\begin{abstract}
Quantum field theories (QFTs) at finite densities of matter generically
involve complex actions. Standard Monte-Carlo simulations based upon
importance sampling, which have been producing quantitative first principle
results in particle physics for almost fourty years, cannot be
applied in this case. Various strategies to overcome this so-called
{\it sign problem} or {\it complex action problem} were proposed 
during the last thirty years. We here review the sign problem in
lattice field theories, focussing on two more recent methods:
Dualization to world-line type of representations and the density-of-states
approach. 

\end{abstract}

\ccode{PACS numbers: 11.15.Ha,12.38.Gc}

\section{Sign problem essentials} 

Since their infancy, Monte Carlo simulations on
space-time lattices have evolved into a powerful quantitative tool for
ab-initio calculations in quantum field theory. However, Monte Carlo
methods face major problems when the action $S$ becomes complex and
the Boltzmann factor $e^{-S}$ cannot be used as a weight in a
stochastic process. Examples for this so-called {\it sign problem} or
{\it complex action problem} (we use the two terms synonymously) 
are theories with chemical potential or models with a vacuum
term. Different strategies were explored, and the reviews at the yearly
lattice conferences 
\cite{Borsanyi:2015axp, Sexty:2014dxa,Gattringer:2014nxa,Aarts:2013lcm,Wolff:2010zu,deForcrand:2010ys,Chandrasekharan:2008gp}
summmarize the progress. 

After an introduction to the problem, we focus in this short review on
two recent strategies for dealing with the complex action problem:  
The dual approach and density of states techniques. In the dual
approach the complex action problem is solved completely by exactly
mapping the theory to new variables, the partition sum of which has
only real and positive contributions, such that Monte Carlo sampling
is feasible. The method is powerful and elegant, but it is not yet
clear for which classes of models real and positive dual
representations can be found.  The density of states approach, on the
other hand, is a generally applicable strategy, where the challenge is
to get under control the numerical accuracy needed for reliable results.

\subsection{What is the sign problem?}
The aim in finite temperature quantum field theory is to calculate expectation values
\be
\langle A \rangle = \frac{1}{Z} \, \Tr \, \Bigr[ A \, \exp (- 
  H/T)  \, \Bigr] , \hbo Z = \Tr \, \exp (- H/T )  \, , 
\label{eq:k1}
\en 
where $H$ is the Hamiltonian, $T $ the temperature, $Z$
the partition function and $A$ an operator representing an
observable, i.e., $[H,A]=0$. For studying such quantum systems
with Monte-Carlo methods they are mapped to a path integral, 
\be
Z = \sum _{c_1 \ldots c_n}
\langle c_1 \vert \mathrm{e}^{-aH} \vert c_2 \rangle \, \ldots \, 
\langle c_{n-1} \vert \mathrm{e}^{-aH} \vert c_1 \rangle \; =: \;
\sum_{\{c\}} P(c) \; .
\label{eq:k2}
\en
$P(c)$ is referred to as the {\it probabilistic weight} of
the (classical) {\it configurations} $c = c_1 \ldots c_n$, and the lattice spacing $a=\frac{1}{n \, T}$ is the regulator we introduce.  
Similarly
\be
\Tr \, \Bigr[ A \, \exp (-  H/T)  \, \Bigr] \; = \;
\sum_{\{c\}} A(c) \, P(c) \; . 
\label{eq:k3}
\en
{\bf If} $P(c)$ is (semi-)positive definite, Monte-Carlo simulations
generate $m$ configurations $c_k$, $k=1\ldots m$ using {\it importance 
sampling} with the weight $P(c)$. Expectation values and the error are
estimated by 
\be
\langle A \rangle \approx \frac{1}{m} \sum _{k=1}^m A(c_k) , \hbo
\hbox{err}_A \approx \sqrt{\frac{ 2 \tau +1 }{m-1}}\,  \sigma _A \; ,
\label{eq:k4}
\en 
where $\tau $ is the auto-correlation time, and $\sigma _A$ the
standard deviation of the $A(c_k)$. 

For studying the grand-canonical ensemble we introduce a
chemical potential $\mu $ and generalise to
$H \, = \, H_0  + \mu \, N$
where $N$ is the particle number operator  with  $[H,N]=0$. The key observation is that
when repeating the steps to derive the path integral, it generically turns out that, even for
bosonic theories, the ``weights'' $P(c)$ are complex numbers. We
illustrate this here for a (1-dimensional) quantum mechanical system,
$
H_0 = \frac{1}{2} \hat{p}^2 + V(x), \; N = \hat{p},
$
where $\hat{p}$ is the momentum operator. Using momentum $\vert p \rangle $
and coordinate $\vert x \rangle $ eigenstates, we find:
\bea 
&& \langle x_n \vert \e^{-aH} \vert x_{n+1} \rangle =
\int _{p_n} \langle x_n \vert \e^{-a V(x_n)} \vert p _n \rangle \,
\langle p_n \vert x_{n+1} \rangle \; \exp \Big(- \frac{a}{2} p_n^2 - a
\mu \, p_n \Big) \; =
\label{eq:k7} \\
&& \int _{p_n} \!\!\! \! \e^{-a V(x_n) } \; \e^{ - \frac{a}{2} p_n^2 - a
  \mu \, p_n } \; \e^{i p_n (x_{n+1} - x_n) }
\; \propto \; \exp \!\left(\! - \frac{a}{2}  \left[ \frac{ x_{n+1}\! -\!
    x_n}{a}  +  i \mu \right] ^2  -  a \, V(x_n) \! \right)\! .  
\nonumber 
\ena
In the limit $a\to 0$, the partition function can be written as an
ensemble average over closed ``world lines'' parameterised by $x(\tau
)$ with $x(0)=x(L)$, $L=1/T$:
\be
Z \propto \int {\cal D}x \; \exp \left( - \int _0^L d\tau \left[
  \frac{1}{2} \left( \dot{x} + i \mu \right)^2  \, - \, V(x(\tau))
  \right] \, \right) .
\label{eq:k8}
\en
The ``string-inspired'' representation of quantum systems in terms of
world lines has been used to calculate $n$-point amplitudes in
perturbative quantum field theory (for a review
see~\cite{Schubert:2001he}). The world-line formalism also is a
viable numerical tool~\cite{Gies:2001zp}, which was 
applied to study the quantum interaction of
vortices~\cite{Langfeld:2002vy} or the Casimir
effect~\cite{Gies:2003cv}.

Here we make two important observations: (i) Although Eq.~(\ref{eq:k1}) shows that the partition
function $Z$ is real and positive, in the form (\ref{eq:k2}) $Z$ only becomes real  
in the ensemble
average, since $P(c)$ in general can be complex. Complex $P(c)$ renders   the standard
Monte-Carlo method impossible since we loose the   probability
interpretation.  (ii) Whether $P(c)$ is real or complex is
representation  dependent. If we would, e.g., choose the exact
eigenstates of the   Hamiltonian $H$, $P(c)$ would be real and positive (note,
however, that knowing these eigenstates means that we already solved
the problem exactly).

Since the mere existence of the sign problem seems artificial, a
proper definition of the problem and of what we would consider a
solution is essential. We here follow closely the discussion by Troyer
and Wiese~\cite{Troyer:2004ge}: 
\bii
\item A quantum system is defined to suffer from a sign problem if
  there occur negative (or complex) weights $P(c)$ in the classical
  representation.
\item An algorithm for the stochastic estimate of an expectation value
  such as $\langle A \rangle $ is of {\it polynomial complexity}
  if the computational time $t(\epsilon,V,T)$ needed to
  achieve a statistical error $\epsilon = \mathrm{err}_A/\langle A
  \rangle $ scales polynomial with the system size $V$, i.e.,
  there exist integers $n$ and $\nu $ (and a finite constant $\kappa
  $) such that 
$
  t(\epsilon,V,T) < \kappa \, \epsilon ^{-2} \, \frac{V^n}{T^\nu} \; . 
$
\item For a quantum system suffering from a sign problem for $A$,
  for which there exists a polynomial complexity algorithm for a
  related classical system, we call this algorithm a {\it solution of
    the sign problem} for the calculation of $\langle A \rangle $.
\eii

\subsection{Why does the re-weighting algorithm not solve sign problems?} 
\label{sec:why} 

In the eighties {\it re-weighting} was proposed for quantum systems
where the weight $P(c)$ of the
corresponding classical formulation is complex (or not  strictly positive):
\be
P(c) \; = \; \exp ( i \varphi (c) ) \; \vert P(c) \vert \; , \hbo
\varphi (c)  \not=0 \; \; \; \hbox{ for some } \; c.
\label{eq:k11} 
\en
If Monte-Carlo configurations $c^\prime $ are generated with respect
to $\vert P(c) \vert $, the expectation value in (\ref{eq:k1}) can
formally be written in terms of estimators with respect to configurations
$c^\prime$:
$
\langle A \rangle \; = \; \langle A(c^\prime) \;  \exp ( i \varphi
  (c^\prime)   ) \rangle_R / \langle \exp ( i
  \varphi (c^\prime )  )  \rangle _R  \, . 
$
The key observation is that the expectation values $\langle
\ldots \rangle _R$ {\it cannot} be estimated in polynomial time
for a given relative error $\epsilon $. Hence, the re-weighting
algorithm fails one of the critera for qualifying as
solution. We illustrate this for the phase factor
expectation value. Following, e.g.~\cite{Troyer:2004ge}, we write this
expectation value as ratio of two partition functions,
\be
\Bigl\langle \exp ( i  \varphi (c^\prime )  )   \Bigr\rangle _R \; =
\; \frac{Z}{Z^\prime } \; , \hbo Z \;= \; \sum _c p(c) , \; \; \;
Z^{\prime } \; = \; \sum _c \vert p(c) \vert \; .
\label{eq:k16}
\en
The quantum systems related to $Z$ and $Z^\prime $ possess different
free energy densities, denoted as $f$ and $f^\prime $. Hence, we find
\be
\Bigl\langle \exp ( i  \varphi (c^\prime )  )   \Bigr\rangle _R \; =
\; \exp \left( - \frac{ \Delta f }{T} \, V \right) \; , \hbo
\Delta f \equiv f - f^\prime  \; . 
\label{eq:k17}
\en
The triangle inequality implies  
$
\Bigl\langle \exp ( i  \varphi (c^\prime )  )  \Bigr\rangle _R \;
\le \; 1
$,
giving rise to $ \Delta f \, \ge \, 0$. 
Thus, we find that $Q :=\langle \exp ( i \varphi ) \rangle $ is
exponentially suppressed with the volume. In an actual Monte-Carlo
simulation, a significant portion of the configurations produce
contributions  $\exp ( i  \varphi (c^\prime )  ) $ of order
$1$. Hence, the error for the phase factor is only suppressed by the
Monte-Carlo ``run-time'' $m$ leaving us with 
\be 
\hbox{err}_Q \; = \; \frac{\sigma}{\sqrt{m}}, \hbo
\frac{ \hbox{err}_Q }{Q} \; = \;
\frac{\sigma}{\sqrt{m}} \, 
\exp \left( \frac{ \Delta f }{T} \, V \right) \; . 
\label{eq:k18}
\en 
As expected, the re-weighting is {\it not} of polynomial complexity. 

In the remainder of this paper, we focus on two methods that bear the
potential to be of polynomial complexity. (i) The exact reformulation of a model to a 
dual formulation in terms of new variables such that all $P(c)$  are real and positive (see Section~\ref{sec:dual}).
Moreover, the dual version quite often allows for
very efficient simulations using flux or worm algorithms. (ii) The
complexity of the so-called density-of-states method (see
Section~\ref{sec:LLR}) is not yet clear: 
however, this approach is not entirely stochastic - it has an element
of direct integration, and secondly, the method features an
exponential error suppression, which might help to counterbalance
any difference in the free energy densities such as in (\ref{eq:k18}). 

\section{Dualization and flux simulations \label{sec:dual}}	

As already mentioned, the dual approach consists of a mapping of the
theory to new variables (dual variables) such that the partition sum
has only real and positive  contributions. Monte Carlo sampling then
is possible in terms of the dual variables. The dual variables
approach is powerful and elegant, but it is a priori unclear for which
class of theories a real and positive dual representation is possible.  

However, some of the structure of the dual variables is well
understood by now and the toolbox for dual mappings is growing
continuously. In general one trades the conventional representation in
terms of classical fields in a path integral for new integer valued
variables which, however, are subject to constraints such that they
give rise to a geometrical interpretation in terms of world-lines and
world-sheets. 

The dual variables for matter can be viewed as closed loops on the
links of the lattice. The flux along the loops is unbounded for
bosons, while it can only be 0 or $1$ for each element of fermion
flux. The dual variables for gauge fields are surfaces, which either
are closed surfaces  or surfaces bounded by matter flux. Alternatively
the surfaces can be viewed as being built from cycles of flux around
elementary plaquettes, and  again the flux around these cycles can be
from all integers. We stress at this point that the dual
representation of a lattice field theory is not unique. For some
systems several different dual representations are known and it is an
interesting open question how these are related and how the underlying
symmetries of a theory in the conventional representation manifest
themselves in the dual form. 

In this section of our review, we discuss dual variables by first
presenting the general idea of the dual approach for a simple  charged
scalar field (relativistic Bose gas). Subsequently we discuss the
generalization to systems with abelian gauge fields  and conclude the
presentation of the dual approach with addressing open challenges. 

\subsection{A prototype example: The dual form of a charged scalar field}
\label{dual1}

A theory which is well suited for presenting the idea of the dual
approach is the charged scalar field which is described by the lattice
action 
\begin{equation}
S \; = \; \sum_x \!\left( \eta |\phi_x|^2 + \lambda |\phi_x|^4 - 
\sum_{\nu = 1}^4 \left[ e^{\mu \, \delta_{\nu,4} } \phi_x^\star
  \phi_{x+\widehat{\nu}}  \, + \, e^{-\mu \, \delta_{\nu,4} }
  \phi_x^\star \phi_{x-\widehat{\nu}}  \right]  \right) . 
\label{action}
\end{equation}
The conventional degrees of freedom are the complex valued fields
$\phi_x \in \mathds{C}$ on the sites $x$ of a 4-dimensional lattice
with  periodic boundary conditions. We use $\eta = m^2 + 8$, where $m$
is the mass of the field. By $\lambda$ we denote the quartic coupling
and  $\mu$ is the chemical potential, which gives a different weight
to forward and backward hopping in time direction ($\nu = 4$).  It is
obvious that for finite $\mu$ the action has a non-vanishing imaginary
part and in the conventional  representation the model has a complex
action problem.  The partition sum $Z = \int D[\phi] e^{-S}$ is
obtained by integrating the Boltzmann factor $e^{-S}$ with  the
product measure $\int D[\phi] = \prod_x \int_\mathds{C} \frac{d
  \phi_x}{2\pi}$.  

For obtaining the dual representation we now follow$\,$
\cite{Gattringer:2012df,Gattringer:2012ap} $\!\!$:  The contribution
from the nearest neighbor terms of (\ref{action}) to the Boltzmann
factor is written as the product 
\begin{eqnarray}
\hspace*{-7mm}&& \prod_{x,\nu} \exp\!\left( e^{\mu \, \delta_{\nu,4} }
\phi_x^\star \phi_{x+\widehat{\nu}}\right)   \exp\!\left( 
e^{-\mu \, \delta_{\nu,4} } \phi_x \phi_{x+\widehat{\nu}}^\star \right)  =   
\label{nnterm}  \\
\hspace*{-7mm}&& 
\sum_{\{ n, \overline{n}\}} \!\!
\left( \prod_{x,\nu}\! \frac{1}{n_{x,\nu}! \, \overline{n}_{x,\nu}!} \right) \!
\left(\! \prod_x e^{\,\mu [ n_{x,4} - \overline{n}_{x,4} ] } \right) \!
\left( \prod_{x,\nu}\! \Big(\phi_x^\star \phi_{x+\widehat{\nu}}\Big)^{n_{x,\nu}} \, 
\Big(\phi_x \phi_{x+\widehat{\nu}}^\star\Big)^{\overline{n}_{x,\nu}} \! \right) \!= 
\nonumber \\
\hspace*{-7mm}&& 
\sum_{\{ n, \overline{n}\}} \!\!
\left( \prod_{x,\nu}\! \frac{1}{n_{x,\nu}! \, \overline{n}_{x,\nu}!} \right) \!
\left(\! \prod_x e^{\,\mu [ n_{x,4} - \overline{n}_{x,4} ] } \,
     {\phi_x^{\,\star}}^{\sum_\nu [ n_{x,\nu} +
         \overline{n}_{x-\widehat{\nu},\nu} ] }  \, {\phi_x}^{\sum_\nu 
[ \overline{n}_{x,\nu} + n_{x-\widehat{\nu},\nu} ] }  \right) \! ,
\nonumber 
\end{eqnarray}
where in the second step we expanded the individual exponentials and
subsequently reorganized the terms in the product. We use the
abbreviation $\sum_{\{n,\overline{n}\}}  =  \prod_{x,\nu} \,
\sum_{n_{x,\nu} = 0}^\infty  \; \sum_{\overline{n}_{x,\nu} =
  0}^\infty$ for denoting the sum over all configurations of the
expansion indices $n_{x,\nu}, \overline{n}_{x,\nu}  \in
\mathds{N}_0$. In the form (\ref{nnterm}) one can now integrate the
contribution from the nearest neighbor terms with the on-site measures
$\prod_x \int_\mathds{C} \frac{d \phi_x}{2\pi} \, e^{- \eta |\phi_x|^2
  - \lambda |\phi_x|^4}$.  

An important step in the dualization is to separate the degrees of
freedom (dof.) which reflect the symmetry of the theory from all other
dof., since integrating out the dof.\ related to symmetries will
generate the constraints for the dual variables. For our example the
symmetry  is a global U(1) rotation of the fields, $\phi_x \rightarrow
e^{i\alpha} \phi_x$ which leaves the action (\ref{action}) and the
measure  $\int \!  D[\phi]$ invariant. For the terms that remain in
the effective measure $\prod_x \int_\mathds{C} \frac{d \phi_x}{2\pi}
\,  e^{- \eta |\phi_x|^2 - \lambda |\phi_x|^4}$ this is even a local
symmetry. To separate the symmetry from the radial dofs.\ we write the
field variables in polar coordinates $\phi_x = r_x\,e^{i \theta_x}$,
and obtain for the partition sum 
\begin{eqnarray}
\hspace*{-3mm} Z & = & \sum_{\{ n, \overline{n}\}}  
\left( \prod_{x,\nu}\! \frac{1}{n_{x,\nu}! \, \overline{n}_{x,\nu}!} \right) \!\!
\left(\! \prod_x \int_{-\pi}^\pi \frac{d\theta_x}{2\pi}
e^{-i\theta_x \sum_\nu [ n_{x,\nu}  - \overline{n}_{x,\nu} - 
( n_{x-\widehat{\nu},\nu}   - \overline{n}_{x-\widehat{\nu},\nu}) ] }\! \right)  \nonumber \\
&\times &\!
\left(\! \prod_x e^{\,\mu [ n_{x,4} - \overline{n}_{x,4} ] }
\int_{0}^\infty\!\!\! dr_x \; r_x^{1 + \sum_\nu [ n_{x,\nu} + n_{x-\widehat{\nu},\nu} 
    + \overline{n}_{x,\nu} + \overline{n}_{x-\widehat{\nu},\nu} ] }  \;
e^{-\eta r_x^2 - \lambda r_x^4} \right)\! . 
\end{eqnarray}
The integrals over the phases give rise to Kronecker deltas which
implement the constraints. To make more evident the flux nature of the 
dual variables we finalize the dualization by a change of the discrete
variables and switch to new variables 
$k_{x,\nu} \in \mathds{Z}$ and $l_{x,\nu} \in \mathds{N}_0$, 
which are related to $n_{x,\nu}, \overline{n}_{x,\nu}$ by
\begin{equation}
n_{x,\nu} - \overline{n}_{x,\nu} = k_{x,\nu} \qquad
\mbox{and} \qquad  n_{x,\nu} + \overline{n}_{x,\nu} = |k_{x,\nu}| + 2 l_{x,\nu} \; ,
\end{equation} 
and constitute the set of dual variables used in the final form of the
partition sum  
\begin{equation}
  Z  =  \sum_{\{ k, l\}}  \left( \prod_{x,\nu}\!
  \frac{1}{(|k_{x,\nu}| + l_{x,\nu})! \, l_{x,\nu}!} \right)
  \left(\prod_x W(s_x) \right) e^{\mu \sum_x k_{x,4}} 
\prod_x \delta\left( \nabla_{\!\nu} \,  k_{x,\nu} \right) , 
\label{Zfinal}
\end{equation}
where $W(n) = \int_0^\infty dr e^{-\eta r^2 - \lambda r^4} \, r^{n+1}$, 
$s_x = \sum_\nu \big[ |k_{x,\nu}| +  |k_{x-\widehat{\nu},\nu}| + 2(
  l_{x,\nu} + l_{x-\widehat{\nu},\nu}) \big]$, 
and we denote the discrete divergence as $\nabla_{\! \nu} \,
k_{x,\nu} = \sum_\nu \big[ k_{x,\nu}  -  k_{x-\widehat{\nu},\nu}
  \big]$. For a numerical  simulation the $W(n)$ can easily be
pre-computed numerically and stored.  

In the final form (\ref{Zfinal}) the partition function is written as
a sum over all configurations of the dual variables  $k_{x,\nu} \in
\mathds{Z}$ and $l_{x,\nu} \in \mathds{N}_0$. The dual variables
$k_{x,\nu}$ are subject to constraints which were generated when  
integrating out the phases $\theta_x$ which correspond to the global U(1)
symmetry of the original theory and even a local symmetry of the  
on-site measure $\prod_x \int_\mathds{C} \frac{d \phi_x}{2\pi} \, 
e^{- \eta |\phi_x|^2 - \lambda |\phi_x|^4}$. The constraints
$\nabla_{\!\nu} \,  k_{x,\nu} = 0, \forall x$ imply vanishing
divergence for $k_{x,\nu}$, i.e., the $k_{x,\nu}$ have the
interpretation of conserved flux. The dual variables $l_{x,\nu}$ are
not subject to constraints and via $s_x$  enter in the weight factors
$W(s_x)$ from integrating the radial degrees of freedom. It is obvious
that also for finite $\mu$ the  dual partition sum (\ref{Zfinal}) has
only real and positive contributions and the complex action problem is
solved completely.   

In the dual form the particle number has a beautiful geometrical
interpretation: We already noted that the discrete flux $k_{x,\nu}$ is
constrained to vanishing divergence $\nabla_{\!\nu} \,  k_{x,\nu}  =
0$, and the admissible configurations of $k_{x,\nu}$ are closed flux
lines. This gives rise to a simple interpretation of the particle
number $\cal{N}$  which in the grand canonical ensemble couples in the
form $e^{\mu \beta \cal{N}}$, where $\beta$ is the inverse
temperature.  In the dual representation $\mu$ couples to  $\sum_x
k_{x,4}$, i.e., the total flux in time direction  (= the
4-direction). Since the $k$-flux is conserved this sum equals to $N_t
\, w[k]$, where $N_t$ is the temporal extent of the lattice   and
$w[k]$ is the winding number of the flux around the compact time
direction. Since in lattice units $\beta = N_t$, we identify $\cal{N}
= $ $w[k]$, i.e., the particle number  is given by the temporal
winding number of the flux.  

So far we only discussed the transformation of the partition sum to
the dual representation. However, for addressing physical questions we
also need to identify observables in terms of the dual variables. The
simplest way to obtain vacuum expectation values of observables is by
derivatives  of $\ln Z$ with respect to the couplings of the theory. A
derivative with respect to $\mu \beta$ will give rise to the
expectation value  $\langle \cal{N} \rangle$, where in the dual
representation the particle number $\cal{N}$  is given by the winding
number $w[k]$ of $k$-flux as discussed above.  A derivative with
respect to $\eta$ will give rise to the expectation value $\langle
\sum_x  | \phi_x |^2 \rangle$,  which in the dual representation is   
a sum of ratios of weight factors, $| \phi_x |^2 \sim \sum_x W(s_x
+2)/W(s_x)$. In a similar way also the corresponding susceptibilities
can be expressed by sums of dual weights. 

More generally, one can also make the couplings depend on the lattice point and take derivatives with respect to these local couplings to generate 
$n$-point functions. The dualization goes through without major changes and after taking the respective derivatives all local couplings are set to 
the desired global value. The $n$-point functions are then expressed in terms of corelators of the dual variables. In some cases an even more 
general approach is possible, where a more general dualization in the presence of source terms can be found. These source terms typically 
give rise to a larger space of dual variables, e.g., by introducing new dual elements such as open strings with endpoints. Again one can perform
derivatives with respect to the sources also in the dual form of the partition sum and finds a representation of the observables in the enlarged dual 
space (see, e.g., \cite{Korzec:2011gh,Gattringer:2012ap,Rindlisbacher:2016zht}).

An important challenge is how to efficiently update the dual systems. We have seen that the dual variables are subject to constraints, which 
give rise to the interpretation of closed loops of flux. Consequently, for a Monte Carlo update the trial configurations one proposes also 
have to obey the constraints. For bosonic theories local changes, such as changing the loop along a plaquette combined with proposing 
globally winding loops are an option, however, usually are rather inefficient. An elegant and often very efficient approach is the
Prokof'ev-Svistunov worm algorithm \cite{Prokof'ev:2001zz}. Here one violates the constraints at a site of the lattice and subsequently propagates 
the defect along links of the lattice thus creating the "worm". Each step of the worm is accepted with a Metropolis decision based on the local 
changes of the weight. The update is concluded once the head of the worm reaches the starting point and all violated constraints are healed.  
Various generalizations of the worm algorithm were discussed in the context of lattice field theories -- see, e.g., 
\cite{Wolff:2008km,Korzec:2011gh,Mercado:2012ue,Mercado:2012yf} and several of the papers cited below.   

With dualization techniques similar to those discussed for our example of the scalar field, successful, 
i.e., real and positive
dual mappings were obtained for several scalar field theories: The O(2) model with chemical potential \cite{Banerjee:2010kc}, 
various variants of $\phi^4$ theories \cite{Wolff:2009ke,Weisz:2010xx,Hogervorst:2011zw,
Siefert:2014ela,Gattringer:2012df,Gattringer:2012ap}, O(N) models 
\cite{Endres:2006zh,Endres:2006xu,Wolff:2009kp,Bruckmann:2015sua,Bruckmann:2015hua,Bruckmann:2015uhd},
CP(N-1) models \cite{Wolff:2010qz,Bruckmann:2015sua} and the SU(2) chiral model \cite{Rindlisbacher:2015xku}. 
An interesting line of work are the papers by the Berlin group, where 
high precision dual simulations were used to revisit the problem of triviality of $\phi^4$ theory 
\cite{Siefert:2014ela,Hogervorst:2011zw,Weisz:2010xx,Wolff:2009ke}.
Among the systems relevant for particle physics are also spin systems with a chemical potential which serve as effective theories 
for finite density QCD 
\cite{Patel:1983sc,DeGrand:1983fk,Karsch:1985cb,Bergner:2013qaa,Langelage:2014vpa,Langelage:2014vpa,
Glesaaen:2015vtp,Bergner:2015rza}. Also for some of these models successful dualizations based on similar techniques as discussed above were 
presented in recent years \cite{Mercado:2011ua,Gattringer:2011gq,Mercado:2012ue,Mercado:2012yf}.

\subsection{Theories with gauge fields}

Let us now switch from scalar field theories to the more involved case of Abelian gauge theories and Abelian gauge-Higgs 
theories\footnote{So far real and positive dual representations were found only for abelian gauge fields.}. 
For the case of U(1) gauge fields the dualization 
with a strategy similar to the one used in the last section is straightforward (we here follow the presentation in \cite{Mercado:2013ola}). 
The dynamical degrees 
of freedom are the link variables $U_{x,\mu} \in \mathds{C}$ living on the links of the lattice. The action is given by 
$S = - \beta \sum_x \sum_{\mu < \nu} \mbox{Re} \, U_{x,\mu} U_{x + \hat\mu, \nu} U_{x + \hat\nu, \mu}^{\; *} U_{x,\nu}^{\; *}$, such that we find for the
partition sum of pure $U(1)$ lattice gauge theory
\begin{equation}
Z  =  \int \!\! D[U] \, e^{-S}  =  \int \!\! D[U] \prod_x \prod_{\mu < \nu} 
e^{ \frac{\beta}{2} \big[ U_{x,\mu} U_{x + \hat\mu, \nu} U_{x + \hat\nu, \mu}^{\; *} U_{x,\nu}^{\; *}  + 
 U_{x,\mu}^{\; *} U_{x + \hat\mu, \nu}^{\; *} U_{x + \hat\nu, \mu}U_{x,\nu} \big] } \; ,
\label{U1step1a}
\end{equation}
where the path integral measure is given by $\int \!\! D[U] = \prod_{x,\mu} \int_{U(1)} dU_{x,\mu}$. 

For the case of U(1) gauge theory, where the contributions to the action are pure phases, 
we can simplify the dualization by using the fact that each factor in the Boltzmann factor has the form of the generating functional 
for modified Bessel functions $I_n$ given by
$e^{\frac{z}{2}[ t + t^{-1}] } =  \sum_{n \in \mathds{Z}}  I_n(z) \, t^n$ where $z,t \in \mathds{C}, t \neq 0$.
We introduce integer valued variables $p_{x,\mu,\nu} \in \mathds{Z}$ and expand each exponent in (\ref{U1step1a}) 
using the generating functional  and find
\begin{eqnarray}
&& Z  =  \int \!\! D[U] \prod_{x} \prod_{\mu < \nu} \sum_{p_{x,\mu,\nu} \in \mathds{Z}} I_{p_{x,\mu,\nu}}(\beta) \, 
\Big(U_{x,\mu} U_{x + \hat\mu, \nu} U_{x + \hat\nu, \mu}^{\; *} U_{x,\nu}^{\; *} \Big)^{p_{x,\mu,\nu}} 
\label{U1step1} \\
&&\!\! = \sum_{\{p\}} \prod_{x,\mu < \nu} \!\!\! I_{p_{x,\mu,\nu}}(\beta) \!\!\!  \prod_{x,\mu < \nu} \int_{U(1)} \!\!\!\!\! \!\!dU_{x,\mu}
 \Big(U_{x,\mu}\Big)^{\!\!- F_{x,\mu}[p] } 
 =  \sum_{\{p\}}  \prod_{x,\mu < \nu} \!\!\! I_{p_{x,\mu,\nu}}(\beta)  \prod_{x, \mu} \delta \Big( F_{x,\mu}[p] \Big),
\nonumber
\end{eqnarray}
with
$F_{x,\mu}[p]  =  \sum_{\rho:\rho < \mu} [ p_{x,\rho,\mu} - p_{x-\hat \rho,\rho,\mu} ] - 
\sum_{\nu:\mu < \nu} [ p_{x,\mu,\nu} - p_{x-\hat \nu,\mu,\nu} ] $.
In the second step of (\ref{U1step1}) we have reorganized the powers of the $U_{x,\mu}$ and $U_{x,\mu}^{\; *} = U_{x,\mu}^{\; -1}$, thus 
collecting the exponents $F_{x,\mu}[p]$ and introduced the sum $\sum_{\{p\}}$ over all configurations of the $p_{x,\mu,\nu}$. In the last step
we have used $\int_{U(1)} dU \, U^n \; = \; \delta(n)$, where $\delta(n)$ denotes the Kronecker delta. Obviously the dualization 
has only real and positive terms. Here the constraints are based on the links of the lattice and require that the combination $F_{x,\mu}[p]$
vanishes at each link. 

Again we have an interesting geometrical interpretation of the dual form of the theory: The plaquette occupation numbers $p_{x,\mu,\nu}$ 
may be viewed as occupation numbers for flux around the contours of the plaquette $(x,\mu,\nu)$, where positive (negative) $p_{x,\mu,\nu}$ 
is for mathematically positive (negative) orientation. The combination $F_{x,\mu}[p]$ thus is the total flux from all plaquettes that contain the
link $(x,\mu)$. The constraint simply implements vanishing total flux at each link of the lattice. If one now for example considers a configuration 
with only plaquette occupation numbers $p_{x,\mu,\nu} = -1, 0$ or $+1$, then the constraints imply that the occupied plaquettes form closed, oriented
surfaces. Since plaquette occupation numbers can simply be added, a general configuration with arbitrary $p_{x,\mu,\nu}$ can be viewed as
a superposition of such closed surfaces. Thus the partition sum (\ref{U1step1}) can also be interpreted as a sum over closed surfaces. The weights 
$I_{p_{x,\mu,\nu}}(\beta)$ then simply take into account the net number $p_{x,\mu,\nu}$ of how often a plaquette $(x,\mu,\nu)$ appears in the
configuration of surfaces. As in the case of the relativistic Bose gas discussed in the previous section we can obtain the relevant observables, i.e.,
the action density and the corresponding susceptibility by derivatives with respect to $\beta$. 
Studies of abelian gauge theories in the dual formulation can be found in various variants (see
\cite{Panero:2005iu,Endres:2006xu,Korzec:2010sh,Korzec:2012fa,Caselle:2014eka,Caselle:2016mqu} for a selection).

In two dimensions the mapping of U(1) gauge theory can easily be generalized by including a topological term, i.e., the addition of 
$i \theta Q$ to the action, where $Q = \frac{1}{2\pi}  \int d^2 x \epsilon_{\mu \nu} F_{\mu \nu}(x)$ is the topological charge. 
In the conventional representation this system now has a complex action problem for $\theta \neq 0$. 
A simple discretization of the U(1) topological charge in 2 dimensions is  
$i \theta Q =  \frac{\theta}{ 2 \pi} \sum_x \mbox{Im}  \; U_{x,1} \,  U_{x+\hat{1},2} \,  U_{x+\hat{2},1}^{\; *} \, U_{x,2}^{\; *}$.
Using this so called field theoretical discretization, 
the combined Boltzmann factor from action and topological charge reads for a single plaquette
\begin{equation}
e^{ \, \eta U_{x} \; + \; \overline{\eta} U_{x}^{\, *}} \; = \; \sum_{p_x \in \mathds{Z}}  
I_{|p_x|}\!\left(2 \sqrt{\eta \overline{\eta}} \, \right) \, \left( \sqrt{\frac{\eta}{\overline{\eta}}} \, \right)^{p_x} \; U_x^{ \,p_x} \; ,
\end{equation}
where $U_{x} = U_{x,1} \,  U_{x+\hat{1},2} \,  U_{x+\hat{2},1}^{\; *} \, U_{x,2}^{\; *}$ and $\eta = \frac{\beta}{2}-\frac{\theta}{4\pi}, 
\bar{\eta} = \frac{\beta}{2}+\frac{\theta}{4\pi}$. In the second step we have already used an expansion formula (see, e.g., \cite{Gattringer:2015nea} for 
a derivation), which generalizes the generating functional for the Bessel functions. Proceeding as before 
one obtains the dualization of 2-d U(1) lattice gauge theory  with a topological term,
\begin{equation}
Z \; = \;  \sum_{\{p\}}  
\prod_x I_{|p_x|}\!\left( 2 \sqrt{\eta \overline{\eta}} \, \right) \, \left( \sqrt{\frac{\eta}{\overline{\eta}}} \, \right)^{p_x} \;  
\delta(p_x-p_{x-\hat{2}}) ~ \delta(p_{x-\hat{1}} - p_x) \; .
\label{zfinal2}
\end{equation}
Again the partition sum is a sum over all configurations of the integer valued plaquette occupation number $p_x \in \mathds{Z}$ subject to 
the constraints implemented by the Kronecker deltas in (\ref{zfinal2}). Also for $\theta$ the partition sum has only real and 
positive contributions, and  (\ref{zfinal2}) is the first successful real and positive dualization for a theory with a vacuum angle (compare 
\cite{Gattringer:2015baa,Kloiber:2014dfa} for some numerical results).

The generalization of the dualization of Abelian gauge theories together with the dualization of the scalar field theory discussed in the previous
subsection to a dualization of Abelian gauge Higgs models is rather straightforward. When gauge fields are coupled to the action of the charged 
scalar given in (\ref{action}), then the nearest neighbor terms are made gauge invariant by multiplying them with the gauge links 
$U_{x,\mu} \in U(1)$. The dual mapping as discussed in Subsection \ref{dual1} goes through essentially unchanged. The only modification is
that now the loops that appear in the dual partition sum (\ref{Zfinal}) are dressed with link variables. These link variables are integrated over
with the gauge action, which is again expanded as in (\ref{U1step1}). Thus the dual variables $k_{x,\nu}$ alter the constraints for the 
plaquette occupation numbers $p_{x,\mu,\nu}$. As a consequence in the gauge-Higgs model the surfaces can have  
boundaries and the link-constraints along the boundaries are saturated by matter flux $k_{x,\nu}$. 

We here give the final result for the dual representation of the U(1) gauge-Higgs model with two flavors of opposite charge. This is an interesting
case because in this electrically neutral system finite density of charges is possible, i.e., a chemical potential can be coupled in a meaningful way.
The fields corresponding to the two charges can have different masses $m$ and $\overline{m}$, different couplings $\lambda$ and 
$\overline{\lambda}$, as well as different chemical potentials $\mu$ and $\overline{\mu}$.
For the two-flavor case we have two sets of dual variables for the matter field, 
the $k_{x,\nu} \in \mathds{Z}, l_{x,\nu} \in \mathds{N}_0$ of (\ref{U1step1})
and another set $\overline{k}_{x,\nu} \in \mathds{Z}, \overline{l}_{x,\nu} \in \mathds{N}_0$ for the second flavor. 
The dual partition function is a sum over 
the configurations of all variables $k_{x,\nu}, l_{x,\nu}, \overline{k}_{x,\nu}, \overline{l}_{x,\nu}, p_{x,\mu,\nu}$,
\begin{eqnarray}
&& \!\!\! Z  =  \!\!\!\!\! \!\! \sum_{\{ k, l, \overline{k}, \overline{l}, p\}}   \prod_{x,\nu}\! \frac{1}{(|k_{x,\nu}| \! +\!  l_{x,\nu})! \, l_{x,\nu}!}   
\frac{1}{(|\overline{k}_{x,\nu}| \! + \! \overline{l}_{x,\nu})! \, \overline{l}_{x,\nu}!}  
 \prod_x W(s_x) \, \overline{W}(\overline{s}_x)   \prod_{x} \prod_{\mu < \nu} \! I_{p_{x,\mu,\nu}}(\beta)  
\nonumber \\
&& \!\!\! e^{\mu \sum_x k_{x,4}} e^{\overline{\mu} \sum_x \overline{k}_{x,4}} 
\prod_x \delta\left( \nabla_{\!\nu}  k_{x,\nu} \right) \delta\left( \nabla_{\!\nu} \overline{k}_{x,\nu} \right) 
  \prod_x \prod_{\nu} \delta \Big( F_{x,\nu}[p] - k_{x,\nu} + \overline{k}_{x,\nu} \Big)\, .  
\end{eqnarray}
All conventions for $s_x$ and $W(s_x)$ are taken over from (\ref{Zfinal})  with $\overline{s}_x$ computed 
from the dual variables $\overline{k}_{x,\nu}, \overline{l}_{x,\nu}$ and $\overline{W}(\overline{s}_x)$ using
$\overline{m}, \overline{\lambda}$. As in the case of the
charged scalar field also here the complex action problem is solved completely, since the chemical potentials $\mu$ and $\overline{\mu}$ 
appear again as coupling to the temporal winding number of $k$ and $\overline{k}$ flux in the real and positive exponential form.

Dual representations for U(1) gauge-Higgs models with chemical potential 
were simulated in \cite{Mercado:2013ola,Schmidt:2015cva}, mainly with the motivation 
of studying condensation phenomena at finite $\mu$, and also the structurally similar case of the $\mathds{Z}_3$ gauge-Higgs system was
analyzed in \cite{Gattringer:2012jt}. Also in two dimensions U(1) gauge-Higgs systems are of interest, since with the topological angle $\theta$ one
has an additional parameter to probe the physics of the system \cite{Gattringer:2015baa,Kloiber:2014dfa}. Concerning the update of the 
gauge degrees of freedom different options were explored \cite{Endres:2006xu,Korzec:2010sh,Mercado:2013yta}. For the more general
case of gauge Higgs systems the fact that the link-constraints of the plaquettes can also be saturated with matter flux allows for an interesting 
generalization of the Prokof'ev-Svistunov worm algorithm \cite{Prokof'ev:2001zz}: After placing a unit of matter flux on 
a randomly chosen link one adds 
plaquettes with occupation numbers $\pm 1$ where for two of the links of the plaquette the constraints are satisfied by matter flux. 
In this way one transports the defects on a link and its endpoints across the lattice in the same way as the usual worm transports
a defect located on sites. Every step of this so-called surface-worm algorithm (SWA) \cite{Mercado:2013yta} is governed by a random decision and 
a Metropolis acceptance step and the SWA was shown to update abelian gauge Higgs systems very efficiently. 

\subsection{Future challenges for the dual approach}

Let us now come to a short discussion of the main open challenges for the dual approach: Non-abelian gauge fields and fermions. Here often 
dualities were studied only in certain limits, in particular for strong coupling and large fermion mass. 

The successful 
dualization we have discussed for the U(1) case is based on expanding the gauge field Boltzmann factor in $\beta = 1/e^2$, i.e., 
it is a strong coupling expansion. Thus also for other gauge groups 
approaches based on the strong coupling expansion can be found in the literature, see, e.g.,\cite{Rossi:1984cv,Karsch:1988zx,
HariDass:2000ca,HariDass:2000tp,HariDass:1999kx,Cherrington:2007ax,
Cherrington:2007is,Cherrington:2008ey,Cherrington:2009ak,Cherrington:2009am,deForcrand:2009dh,Unger:2011it,deForcrand:2014tha}
and some of the work cited below.
Several of these papers 
include only (very) few terms of the strong coupling expansion, while some attempt a complete formal treatment based on the character expansion,
which, however, has manifestly negative weight factors already for pure gauge theories such that a sign problem is actually
introduced by the dualization. It is probably fair to conclude that so far no convincing approach for dualizing non-abelain gauge fields has emerged. 

For fermions the Grassmann nature of the variables in the path integral introduces additional challenges. One can proceed similar
to the charged boson field in Subsection \ref{dual1} and expand the individual Boltzmann factors for the link and on-site terms. At first glance this
looks even simpler than the bosonic case, 
since the series terminate after the linear term due to the nilpotency of the Grassmann variables, thus allowing
only fluxes (or ''occupation numbers'') $0$ and $1$. However, for the subsequent integration the Grassmann variables must be brought into 
their canonical order and the necessary commutations introduce signs for the loops. In addition, also the $\gamma$-matrices 
(or staggered sign factors) introduce signs and phases. Thus in general one is left with signs for 
matter loops summed over  in the partition function. 
Another approach to obtain a loop representation is hopping expansion where with the help of the 
trace-log formula the fermion determinant is written as  
exponential of a sum of loops. This exponential can be expanded, giving a loop representation for the fermionic partition sum, which is
integrated over the gauge fields. However, also here signs from the fermionic character and the $\gamma$-matrices remain. 

A simple way to get rid of the signs is to consider U(1) gauge fields in the strong coupling limit, which reduces the 
theory to a loop model of mesons, i.e., a bosonic loop model without signs (see, e.g., \cite{Salmhofer:1991cc,
Cecile:2007dv,Chandrasekharan:2002ex}). 
Also for other gauge groups loop models were derived in the strong coupling limit and interesting numerical 
results could be obtained, partly by treating the fermions in the large mass limit by including only leading (sometimes resummed) terms
in the hopping expansion \cite{Rossi:1984cv,Karsch:1988zx,DePietri:2007ak,deForcrand:2009dh,Unger:2011it,deForcrand:2014tha,
Rindlisbacher:2015pea,Glesaaen:2015vtp}.

An area with quite some progress are 1+1-dimensional theories with fermions. There the topology of the loops is simpler and
both the signs from the staggered formulation, as well as the signs from the traces over the $\gamma$-matrices of the Wilson 
formulation can be computed in closed form\cite{Stamatescu:1980br}. 
As a result one finds that the massless Schwinger model with
staggered fermions has a real and positive loop representation in the presence of both, a chemical potential or a vacuum term 
\cite{Gattringer:2015nea}. This result can be generalized to a real and positive dual model for a system of 1-dimensional 
nanowires interacting with the 4-dimensional electromagnetic field \cite{Gattringer:2015cxh}. For full QED in four dimensions
a positive dualization was presented for a subset of the fermion loops, so called quasi-planar loops, in \cite{Gattringer:2015pea}.
It is interesting to note that for the massive Schwinger model (except at strong coupling) the signs come back \cite{Gattringer:1999hr}.
Finally, also for 2-dimensional models with 4-fermi interaction several successful dual formulations can be found in the literature
\cite{Gattringer:1998cd,Gattringer:1998ri,Wolff:2008xa,Cecile:2008nb,Endres:2012vd}.

An interesting idea related to some of the techniques presented here is the fermion bag 
approach \cite{Chandrasekharan:2009wc}. 
There the action is split into a free fermion part and a fermion self interaction. For the interaction part the local
Boltzmann factors are expanded and if the corresponding term is activated it locally saturates the Grassmann integrals. 
On the remaining sites, the "fermion bags", one propagates free
fermions with standard methods. The approach was shown to be  
very efficient for several models and interesting physics results were obtained 
\cite{Chandrasekharan:2011mn,Chandrasekharan:2012va,Chandrasekharan:2012fk}.

Let us conclude the section about dual methods with a speculation:
With the growing number of successfully dualized models we begin  to
understand how different symmetries of the conventional formulation
manifest themselves in terms of dual variables. A deeper understanding
of symmetries and dual variables might eventually lead to the
possibility of model building directly in terms of dual
variables. Although many  open questions need to be answered, such as
under which modifications of the dual representation does the
universality class remain the same, and how are different known
dualizations of the same model related to each other, the idea of
formulating models directly in terms  of dual variables is an exciting
perspective for quantum field theories on the lattice. 

\section{The density-of-states approach \label{sec:LLR} }

\subsection{The method \label{sec:method} }

Although the density-of-states (DoS) approach involves a Monte-Carlo
element in one or an another way, it does not fall into the class of
Markov Chain Monte-Carlo simulations. While the DoS $\rho $ enumerates the
amount of configurations for a given action hyper-surface in
configuration space, the partition function is recovered by performing
one-dimensional numerical integrals:
\be 
\rho(E) = \int {\cal D} \phi \;  \delta \Bigl( S[\phi]-E \Bigr) \, ,
\; \; Z(\beta)=\int {\cal D} \phi \;  \e^{\beta S[\phi]} \, = \,
\int \dd E \; \rho(E) \; \e^{\beta E} \, .
\label{eq:k30}
\en
By contrast, conventional Monte-Carlo calculations based on 
importance sampling generate the configurations
contributing to the integral with probability 
\[
P_\beta (E) \; =  \; \rho(E) \; \e^{\beta E}  / Z(\beta ) \,.
\]
In this standard approach, so-called {\it overlap} problems occur if
an observable strongly depends on configurations in an action range
that has low probability according to importance sampling. Key to the
success of DoS techniques is that $\rho (E)$ can be calculated with
good relative precision over the action range of interest. The overlap
problem in this case is avoided by a direct integration of
(\ref{eq:k30}). Wang and Landau~\cite{Wang:2001ab} provided an
efficient algorithm, based upon histograms, for accessing the density
of states in a statistical system with a discrete spectrum. For
systems with continuous spectrum, however, the histogram based method
breaks down even on systems of moderate
size~\cite{Xu:2007aa,Sinha:2007aa}. More sophisticated techniques, as
for instance those proposed by Berg and Neuhaus~\cite{Berg:1992qua}, 
alleviate the problem by using the Wang-Landau method to
compute the weights for a multi-canonical recursion. 

Here, we will focus
on a novel and promising new method, which falls into the class of
Wang-Laudau type approaches and which is called the Logarithmic Linear
Relaxation (LLR) algorithm~\cite{Langfeld:2012ah}. Rather than using
histograms to estimate the slope of the DoS, the method avoids
uncertainties from histogram edge effects and instead employs 
expectation values and a stochastic non-linear equation for this
task: 
\bea
\dlangle W[\phi] \drangle_{k} (a)
&=& \frac{1}{{\cal N}_k} \int {\cal D} \phi \; \theta
_{[E_k,\DE]}(S[\phi]) \; W[\phi] \; \,  \e^{-a S[\phi] } \; , 
\label{eq:k32} \\
{\cal N}_k &=& \int {\cal D} \phi \; \theta _{[E_k,\DE]} \; \, 
\e^{-a S[\phi] } \; = \;  
\int_{E_k}^{E_k+\DE}  \dd E \, \rho (E) \,\mathrm{e}^{-aE} \; , 
\label{eq:k33} 
\ena
where we have used (\ref{eq:k30}) to express ${\cal N}_k$ as an ordinary
integral. We also introduced the modified Heaviside function 
$$ 
\theta _{[E_k,\DE]} (S) \; = \; \left\{ \begin{array}{ll} 
1 & \hbox{for} \; \; \; E_k \leq S \leq E_k + \DE \\ 
0 & \hbox{otherwise . } \end{array} \right. 
$$
Note that $\dlangle W[\phi] \drangle_{k} $ can be estimated by
standard Monte-Carlo techniques. Let us now specialize to the
particular observable $W[\phi] = \Delta E = S[\phi ] - E_k -
\DE/2$, the expectation value of which can be written as an integral
using again (\ref{eq:k30}): 
\be 
\dlangle \Delta E \drangle_{k} (a)
\; = \; \frac{1}{{\cal N}_k} \int_{E_k}^{E_k+\DE}  \dd E \, \rho (E)
\; \Bigl[E - E_k - \frac{\DE}{2} \Bigr] \; \,\mathrm{e}^{-aE} \; .  
\label{eq:k34} 
\en 
At the heart of the LLR algorithm is the stochastic non-linear
equation for determining the parameter $a$~\cite{Langfeld:2012ah}:
\be 
\dlangle \Delta E \drangle_k (a) \; = \; 0 \hbo \Leftrightarrow \hbo 
a \; = \; \frac{ d \; \ln \, \rho }{dE } \Big\vert
_{E=E_k +\frac{\DE }{2}} \; + \; {\cal O}\Bigl(\DE^2\Bigr) \; .
\label{eq:k35}
\en
It was shown~\cite{Langfeld:2015fua} that for sufficiently small $\DE$ there is only one
solution to (\ref{eq:k35}). The solution
$a=a_k$ of the stochastic equation (\ref{eq:k35}) provides the log
derivative of the density-of-states at the midpoint of the action
interval. Finding the solution starts with a standard
Newton-Raphson iteration, which departs from an initial guess $a^{(0)}$
and produces a sequence 
$ 
a^{(0)} \to a^{(1)} \to  \ldots  \to  
a^{(n)} \to  a^{(n+1)} \ldots  
$
using 
\bea 
a^{(n+1)} &=& a^{(n)} \; + \;  \frac{ \dlangle \Delta E \drangle_k
  (a^{(n)})}{ \sigma ^2  (\Delta E; a^{(n)})}  \; ,
\label{eq:k36} \\
\sigma ^2  (\Delta E; a) &=& \dlangle \Delta E^2 \drangle_k (a) \; -
\; \dlangle \Delta E \drangle_k^2 (a) \; = \; - \; \frac{d}{da}
\dlangle \Delta E \drangle_k (a) \;  .
\nonumber 
\ena
For $a^{(n)}$ sufficiently close to the true value $a_k$, we can
approximate $\sigma ^2  (\Delta E; a) \approx \DE^2 /12 $, 
and the Newton-Raphson iteration turns into the fixed point
iteration: 
\be 
a^{(n+1)} \; = \; a^{(n)} \; + \;  \frac{12 }{\DE ^2 }  \; 
\dlangle \Delta E \drangle_k   (a^{(n)}) \; .
 \label{eq:k38}
\en
Note that the approximation for the action fluctuation $\sigma ^2$
does not affect the precision of the solution but rather the rate of
convergence~\cite{Langfeld:2015fua}. In fact, a different
choice for $\sigma ^2  (\Delta E; a)$ was discussed
in~\cite{Gattringer:2015lra} to improve the convergence rate.   
However, the expectation values
$\dlangle \Delta E \drangle_k $ are not known exactly but only
available by means of Monte-Carlo estimators. In practice, after a few
iterations, the uncertainty of the $a^{(n)}$ is dominated by the noise
of the stochastic estimators for $\dlangle \Delta E \drangle_k
$. Since the noise from the iteration mixes with the error of the
iteration, it is not clear a priori that the resulting fluctuations are
normal distributed with a mean around the true solution $a_k$. This
problem, however, has been already solved by Robbins and
Monroe~\cite{robbins1951}. An under-relaxation of the iteration
(\ref{eq:k38}) is essential:
\be
a^{(n+1)} \; = \; a^{(n)} \; + \;  c_n \; \frac{12 }{\DE ^2 }  \; 
\dlangle \Delta E \drangle_k   (a^{(n)}) \; ,
 \label{eq:k39}
 \en
 where the coefficients satisfy
$
\sum^{\infty}_{n=0}c_n = \infty \; \mbox{ and } \;
\sum^{\infty}_{n=0}c^2_n < \infty \, . 
$
It can be shown that if the iteration is truncated at some (large)
$n=n_c$, the corresponding values $a^{(n_c)}$ are normal distributed
with the mean coinciding with the true solution
$a_k$~\cite{robbins1951}. This is indeed how we simulate in practice:
we generate a variety of potential solutions $a^{(n_c)}$ which are
subjected to further calculations of observables the statistical
errors of which we obtain by a standard bootstrap analysis. 

Let us assume now that we have successfully estimated the
log-derivative of the DoS for a variety of different action
intervals. Our initial assumption is that the density of states is a
regular function of the action that can be always approximated in the
finite interval $[E_k,E_k+\DE]$ by a suitable functional expansion.
In this case, we find using Taylor's theorem
\bea
\hspace*{-3mm} \ln  \rho (E) &=& \ln  \rho \left(\!E_k +\frac{\DE}{2} \!\right)  +
\frac{  d \; \ln \, \rho }{dE } \Big\vert _{E=E_k+\DE/2} \!
\left(\!E- E_k -\frac{\DE}{2} \! \right)  +   {\cal O}(\DE^2) . 
\label{eq:k40} 
\ena 
Thereby, for a given action $E$, the integer $k$ is chosen such that 
$
E _k \leq E \leq E_k \, + \, \DE \, , \;
E_k \; = \; E_0 \; + \; k \, \DE \; . 
$
Exponentiating this equation and using (\ref{eq:k35}), it was shown
in~\cite{Langfeld:2015fua} that remarkably 
\bea 
\rho (E) &=& \tilde{\rho } \left( E \right)  
\;  \exp \Bigl\{  {\cal O}(\DE^2) \Bigr\} \; = \; \tilde{\rho } \left(
E \right)   
\Bigl[ 1 \; + \; {\cal O}(\DE^2) \Bigr] \; ,
\label{eq:k41} \\
\tilde{\rho} (E) &=& \rho_0 \left( \prod_{k=1}^{N-1}  \e^{a_k \DE}
\right) \; \e^{a_N (E-E_N) }  \; ,  
\label{eq:k42} 
\ena 
which we will extensively use below. We will observe that $\rho (E)$
spans many orders of magnitude. The 
key observation is that our approximation implements {\it exponential error
  suppression}, meaning that $\rho (E)$ can be approximated with nearly-constant
{\it relative error} despite it may reach over thousands of orders of
magnitude: 
$
1 - \frac{\tilde{\rho}(E)}{\rho (E)} \, = \, {\cal
  O}\left(  \DE^2  \right) . 
\label{eq:k43}
$
Finally, some comments are in order: the above LLR formulation uses
finite size action intervals, which might raise concerns about
ergodicity. In practice we have studied $Z_3$, U(1), SU(2) and SU(3) gauge
theories and have never encountered any ergodicity problem. We also
point out that the ergodicity properties can be easily improved by 
using the {\it replica exchange} method, where one uses overlapping
action intervals and exchanges configurations of neighboring intervals
with the corresponding exchange probability
(see~\cite{Langfeld:2015fua} for details).

\subsection{The SU(2) and SU(3) showcase}

\begin{figure}[t]
\includegraphics[height=5.7cm]{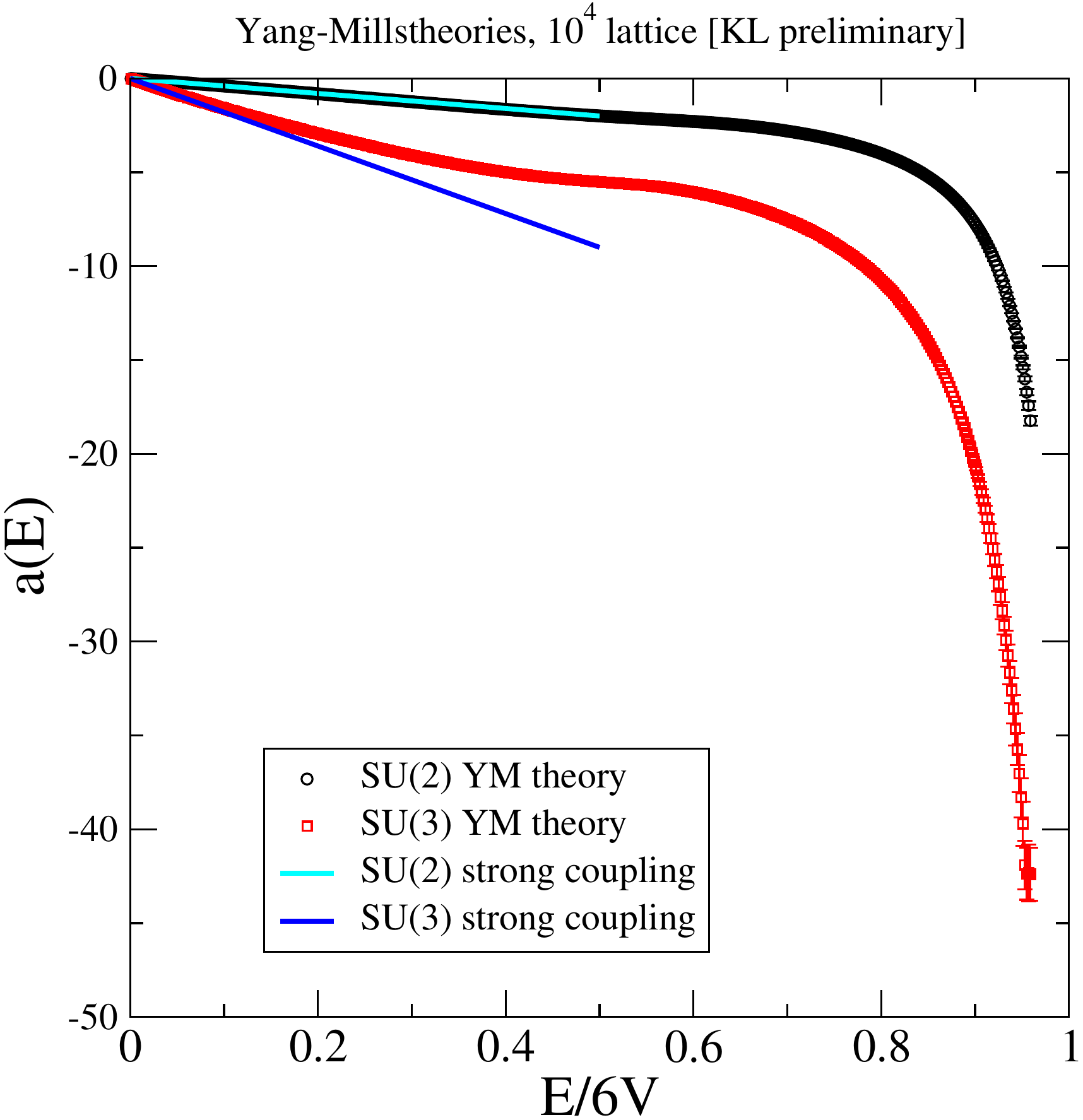} \hspace{0.5cm} 
\includegraphics[height=5.7cm]{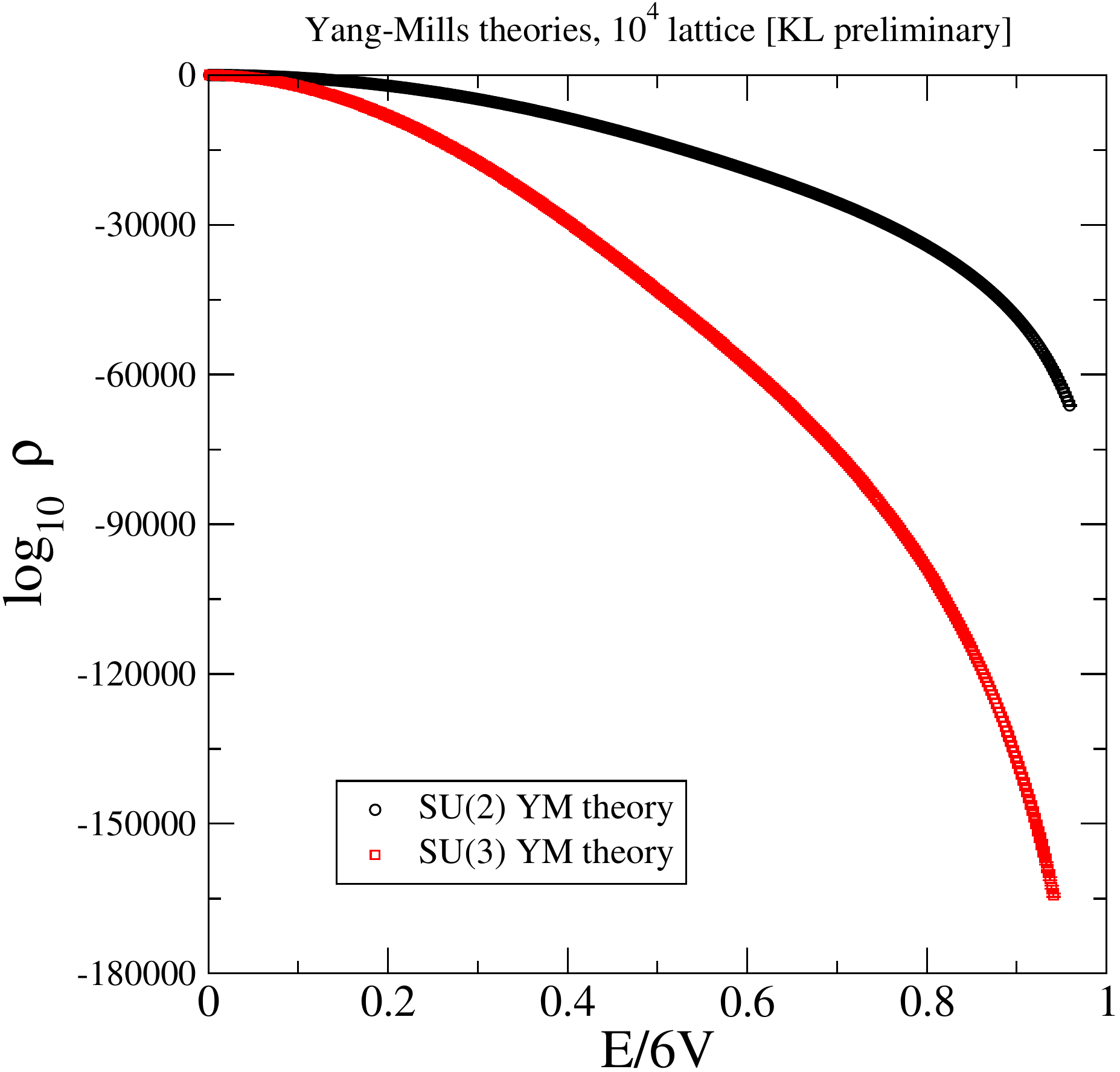}
\caption{Left: The LLR coefficient $a$ as a function of the the action $E$
  in units of its maximal value $6V$, $V=10^4$ is the 4d lattice
  volume. Right: the corresponding density-of-states $\rho (E)$. \label{fig:1}}
\end{figure}
Let us now specialize to the case of $SU(N_c)$, $N_c=2,3$ gauge
theory. The dynamical degrees of freedom are the link variables $U_\mu
(x) \in SU(N_c)$. We will use the Wilson action and show
results for $10^4$ lattices. For each 
action interval, we used $200$ Newton Raphson steps for thermalization
and $600$ Robbins-Monroe under-relaxation steps. We generated $20$
independent candidates for the corresponding LLR coefficient $a$, which
is subsequently used in a bootstrap error analysis. The numerical
findings shown here complement our earlier results 
in~\cite{Langfeld:2012ah}. Our findings for the LLR coefficients for a
SU(2) and a SU(3) theory are shown in Fig.~\ref{fig:1}. 

The probabilistic density generating a lattice configuration within
the action interval $[E, E+ dE]$ consists of an entropy factor, i.e., the density of
states, and the Gibbs factor: 
$
P(E)  =  \rho(E) \, \exp ( \, \beta E ) \, . 
$
Because $E$ is an extensive quantity, $P(E)$ possesses a sharp maximum
(or several of them in case the theory has a first order phase
transition~\cite{Langfeld:2015fua}). The position of the extremal
points can be found by
\be
\frac{dP}{dE} \; = \; P(E) \; \Bigl[ \frac{ d \, \ln \, \rho }{dE} \;
+ \beta \, \Bigr] \; = \; P(E) \; \Bigl[ a(E) + \beta \Bigr] \; = \; 0
\; . 
\label{eq:ky2}
\en 
If the LLR coefficients $a(E)$ are monotonic as a function of $E$,
$P(E)$ has a single maximum for all $\beta $ and does not have a first
order phase transition (for the given aspect ratio of spatial and time-like
lattice size for the simulation). For small $\beta $, the action,
i.e., the plaquette expectation value can be calculated using 
Taylor-expansion with respect to $\beta $ - the so-called strong
coupling expansion: 
\be 
\langle S[U] \rangle \; = \; s_0 \, \beta \; + \; {\cal O} (\beta ^2) \; 
, \;\;\; s_0 = \frac{1}{4} \; \; \hbox{for SU(2)} \; , \; \; \; 
s_0 = \frac{1}{18} \; \; \hbox{for SU(3)} . 
\label{eq:ky3}
\en 
For small $\beta $ only coefficients $a(E)$ close to zero are
relevant in the large volume limit because of (\ref{eq:ky2}). 
Here we approximate to leading order 
$$ 
a(E) \; = \; a_0 \; E \; + \; {\cal O}(E^2) \; , \hbo \rho (E) \; = \;
\rho _0 \; \exp \left( \frac{a_0}{2} E^2 \; + \;  {\cal O}(E^3) \,
\right) \; . 
$$
With this density of states, the action expectation value becomes
$$ 
\langle S[U] \rangle \; = \; \frac{ \int dE \; \rho(E)  \; \exp ( \beta E) \; E }{\int dE
  \; \rho(E) \; \exp ( \beta E) }  \; = \; - \frac{\beta }{a_0} \; . 
$$
Comparing this with the strong coupling result in (\ref{eq:ky3}), we
find at small $E$ that 
\be 
a(E) = - 4 E + {\cal O}(E^2) \; \; \; \hbox{for SU(2)} , \hbo 
a(E) = - 18 E + {\cal O}(E^2) \; \; \; \hbox{for SU(3)}  \; . 
\label{eq:ky4}
\en 
The leading order for $a(E)$ at small values for $E$ is also shown in
Fig.~\ref{fig:1}. Once the coefficients $a(E)$ are known, the
density of states $\rho(E)$ can be estimated with the help of
(\ref{eq:k41}). For each action interval $[E,E+\DE]$, we have
generated $20$ independent candidates for $a(E)$ after $800$
Robbins-Monro iterations. The density and its error are then obtained
by a standard bootstrap analysis. The result is also shown in
Fig.~\ref{fig:1} (right panel). Note the logarithmic scale (base
10). By virtue of the exponential error suppression, we can calculate
the density of states over more than $10^6$ orders of magnitude
with good relative error.

\subsection{The LLR method for theories with a sign problem}

The density-of-states method can be  generalized to
calculate high precision observables other than those depending on
the action. For instance in~\cite{Langfeld:2013xbf} the probability
distribution of the Polyakov line was calculated with extreme
precision for a study of 2-color QCD at finite densities of heavy
quarks. Here we focus on a generalization of the LLR method that
will allow us to simulate theories with a sign problem.

The partition sum of such a theory and its {\it
  phase-quenched} counterpart are  
\be 
Z(\mu ) \; = \; \int {\cal D} \phi \;  \e ^{S_R[\Phi ](\mu )} \; \exp
\Bigr( i S_I[\phi ](\mu) \Bigl)  \; , \; \; \; 
Z_\mathrm{mod} (\mu ) \; = \; \int {\cal D} \phi \; 
 \e ^{S_R[\Phi ](\mu )} \; ,  
\label{eq:k50}
\en 
where $\mu $ is the chemical potential. 
With the help of the phase factor expectation value $Q(\mu)$, we 
trivially cast (\ref{eq:k50}) into
\be
Z(\mu ) \; = \; Q(\mu) \; Z_\mathrm{mod} (\mu ) \; , \; \; 
Q(\mu) \; = \; \frac{Z(\mu)}{Z_\mathrm{mod} (\mu )} \; = \;
\Bigl\langle \exp \Bigl( i S_I[\phi ](\mu) \Bigr) \Bigr\rangle
_\mathrm{mod} .  
\label{eq:k52}
\en
Since observables of the phase-quenched theory are accessible with
standard Monte-Carlo simulations, this implies that the solution of
the sign-problem is relegated to the calculation of $Q(\mu)$. 
The LLR approach to calculate
$Q(\mu)$ starts  with the definition of the density-of-states for the
imaginary part of the action~\cite{Langfeld:2014nta}:
\be 
\rho (s) \; = \; N \; \int {\cal D} \phi \; \delta \Bigl( s \, - \, 
S_I[\phi](\mu) \, \Bigr) \;  \e ^{S_R[\Phi ](\mu )} \; , \; \; 
Q(\mu ) \; = \; \frac{ \int ds \; \rho (s) \; \exp (is ) }{ \; 
\int ds \; \rho (s) } \; . 
\label{eq:k53}
\en
Such a generalized density of states was first introduced by 
Gocksch~\cite{Gocksch:1988iz} to address the phase factor of the quark
determinant of finite density QCD, or for studies of the theta angle
dependence in spin systems in~\cite{Azcoiti:2002nq,Azcoiti:2011ei}.
The theoretical framework of the LLR method as discussed in
Subsection~\ref{sec:method} can be transferred to the case
(\ref{eq:k53}): the intervals now discretise the imaginary part
of the action $S_I$, and after the estimate of the LLR coefficients
$a_k$, the probability distribution $\rho (s)$ can be retrieved. 

\subsection{The $Z_3$ theory at finite densities}

\begin{figure}[t]
\includegraphics[height=5.7cm]{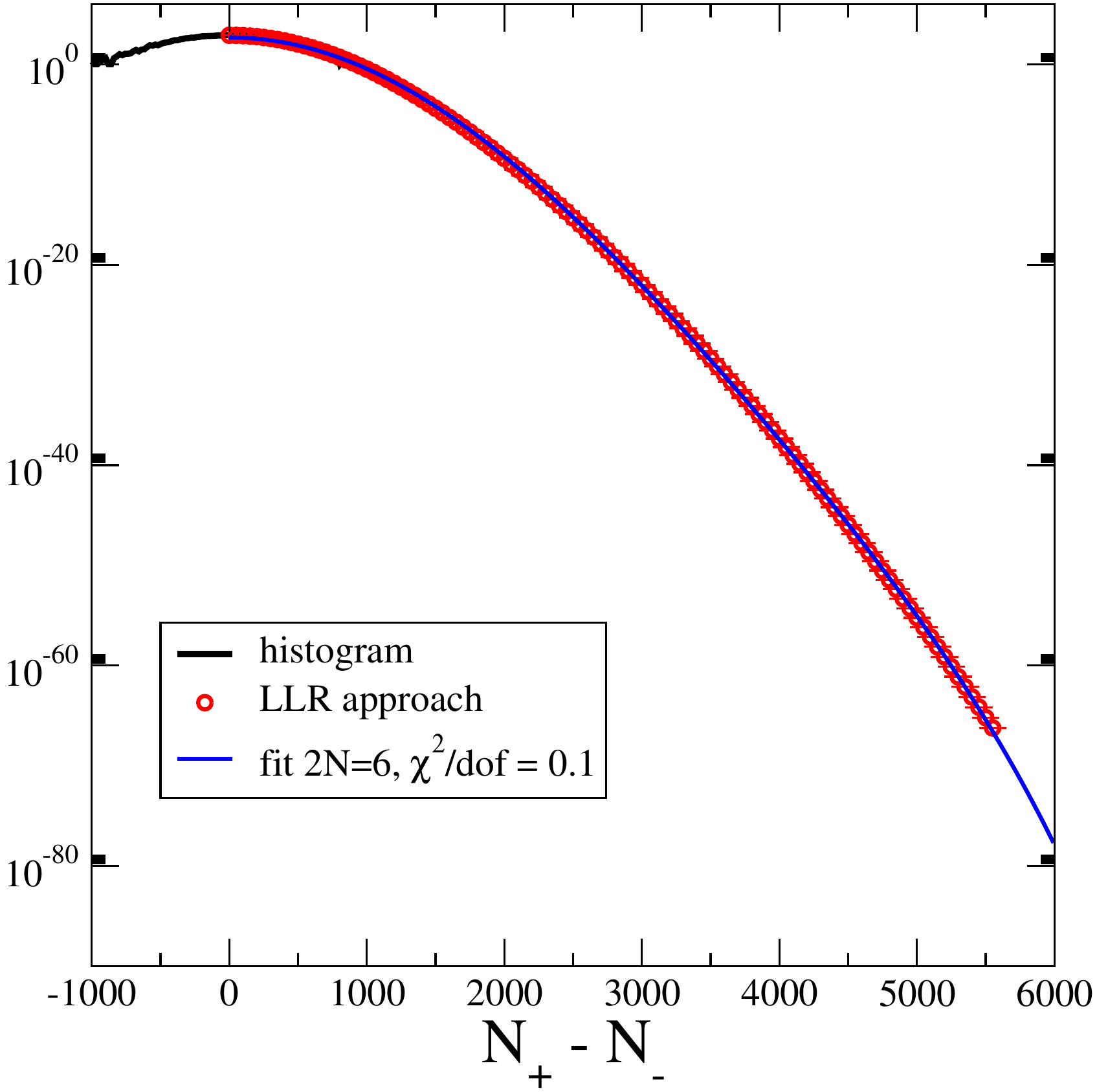} \hspace{0.5cm} 
\includegraphics[height=5.7cm]{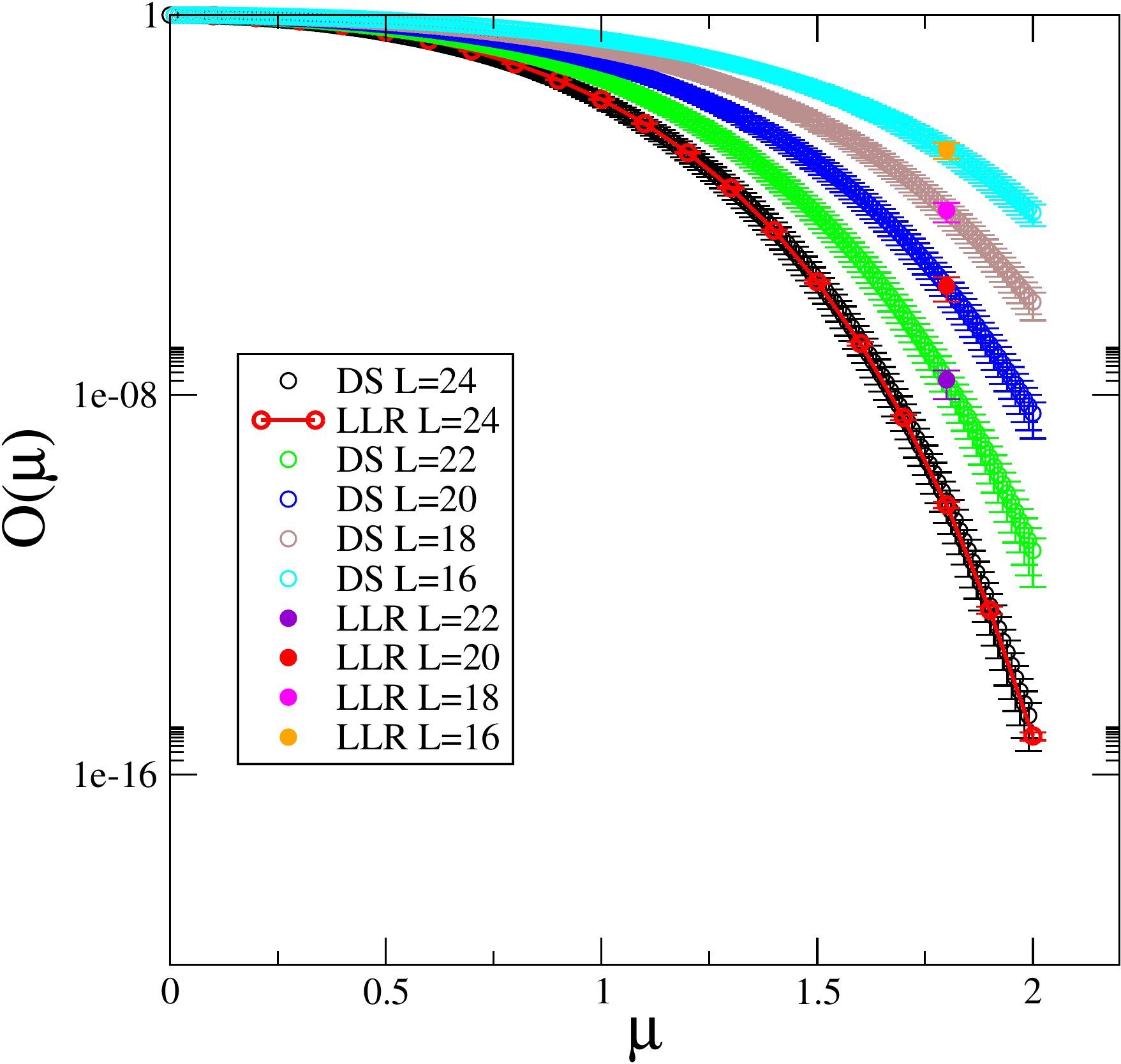}
\caption{Left:  The probability distribution of the centre imbalance
  $\Delta N$ for the $Z_3$ theory on a $24^3$ lattice with $\tau =0.17$
and $\kappa =0.01$. Also shown (red symbols) is the result from the
LLR approach. Right:  The overlap factor as a function of $\mu $; LLR
  results in comparison to the known results from the dual
  theory (figure from~\cite{Langfeld:2014nta}).  \label{fig:2}}
\end{figure}
For a first showcase we briefly review LLR approach to the $Z_3$ spin model at
finite chemical potential $\mu $~\cite{Langfeld:2014nta}. The degrees
of freedom $\phi (x) \in \{1,z,z^\ast \}$, $z=(1+i\sqrt{3})/2$  are
associated with the $N^3$ sites of the  3-dimensional lattice. The
partition function and the action of the system are given by 
\bea
Z(\mu) &=& \sum _{\{\phi\}} \; \exp \Bigl( S[\phi] \Bigr) \; , \; \;
\; 
S[\phi] = \tau \sum _{x,\nu } \phi_x \, \phi^\ast _{x+\nu} + 
\sum_x \, \Bigl( \eta \phi_x + \bar{\eta } \phi^\ast _x \Bigr) \; , 
\label{eq:k60} 
\ena 
with $ \eta = \kappa \, \mathrm{e}^{\mu } $ and $ \bar{\eta }  = \kappa \,
\mathrm{e}^{- \mu } $. This theory has a real dual formulation\cite{DeGrand:1983fk,Patel:1983sc,Mercado:2011ua}, and
can be efficiently simulated with the flux algorithm\cite{Mercado:2012yf}. It is therefore
ideally suited for testing the LLR approach (for a simulation with the related density of states 
FFA (functional fit approach) see \cite{Mercado:2014dva,Gattringer:2015lra,Gattringer:2015eey}). We introduce
\bea 
N_0 &=& \sum_x \delta \Bigl(\phi(x),1 \Bigr) , \; \; \; 
N_z = \sum_x \delta \Bigl(\phi(x),z \Bigr) , \; \; \; 
N_{z^\ast} \; = \;  \sum_x \delta \Bigl(\phi(x),z^\ast \Bigr) , 
\label{eq:k62}
\ena
which correspond to the total number of a particular centre element of
the configuration. It turns out that the imaginary part of the action
is proportional to the {\it centre imbalance} $\Delta N := N_z -
N_{z^\ast}$ on the lattice. Hence, the LLR method targets the
probability distribution of this quantity, and we introduce the
generalized density
\bea 
\rho (n) \; := \; \sum _{\{\phi\}} &&
\delta \Bigl( n, \Delta N[\phi] \Bigr) \; \; 
\exp \Bigl( S[\phi]  
\; + \;   \kappa  \Bigl( 3N_0 [\phi] - V \Bigr) \, \hbox{cosh}(\mu)
 \Bigr) \; . 
\label{eq:k63} 
\ena
The histogram method generates configurations with the Gibbs factor
given by the exponential in (\ref{eq:k63}) and keeps track of $\Delta
N[\phi]$ of each configuration. The result for a $24^3$, $\tau =0.17$
and $\kappa =0.01$ is shown in Fig.~\ref{fig:2}. The histogram
method is riddled by an {\it overlap problem}: configurations with high
$\Delta N$ are hardly generated leading to large relative
uncertainties. The LLR method does solve this problem due to its
inherent exponential error suppression: the density-of-states is
obtained with nearly constant relative errors over $70$ orders of
magnitude. With the density-of-states at hand, the partition function
can be written as a simple sum:  
$
Z(\mu)  =  \sum_{ n} \; \rho ( n) \, \cos ( \sqrt{3} \, \kappa
\,   \hbox{sinh}(\mu) \, n ) . 
$
Although we have reached a very good (relative) precision for $\rho $,
a new challenge arises from the oscillating sum,
which, by virtue of cancellations, results in an exponentially small
result (see (\ref{eq:k17})). Standard interpolation schemes use a
piecewise interpolation of the discrete set of points and seek
convergence by decreasing the spacing $\DE $. It became clear very
early on that this approach lacks the precision to obtain this
signal. We are therefore using an iterative refinement of the
approximation in functional space:
$
\ln \rho (n) =  \lim _{N \to \infty } \sum _{k=1}^N c_k \; f_k(n)
\; , 
$
where $f_k(n)$ are basis functions. The approximation arises from the
truncation of the above sum. For the $Z_3$ spin system,
good results are obtained by using powers of $n$: $f_k(n) =
n^{2k}$. Here we have exploited the symmetry $\rho (-n) = \rho(n)$,
which eliminates odd powers of $n$ from the basis. Fig.~\ref{fig:2},
left panel, shows the method at work: a truncation at $N=3$ already
leaves us with a good representation of the data at the level of
$\chi^2 /\mathrm{dof}=0.1$. 
Once a good functional representation of the data is obtained, the
sums for the overlap factor, 
$$
Q(\mu ) \; = \; \frac{ \sum _n  \rho ( n) \; \cos \Bigl( \sqrt{3} \, \kappa
\,   \hbox{sinh}(\mu) \; n \Bigr) }{ \sum _n  \rho ( n)} \; , 
$$
can be obtained in a (semi-)analytical way. Fig.~\ref{fig:2}, right
panel, shows the result for the overlap factor for several values of
$\mu $~\cite{Langfeld:2014nta}. We find a very good agreement with the
known results from the simulations of the dual real theory.

\section*{Acknowledgements}

We thank F.\ Bruckmann, B.\ Lucini, T.\ Sulejmanpasic, and A.\ Rago  for discussions. KL is supported by 
the  Leverhulme Trust (grant RPG-2014-118) and STFC (grant ST/L000350/1). CG is supported by DFG TRR55, 
and FWF grant I 1452-N27.

\bibliographystyle{ws-ijmpa}
\bibliography{ck_lit}

\begin{thebibliography}{100}
\expandafter\ifx\csname urlstyle\endcsname\relax
  \providecommand{\doi}[1]{doi:\discretionary{}{}{}#1}\else
  \providecommand{\doi}{doi:\discretionary{}{}{}\begingroup
  \urlstyle{rm}\Url}\fi

\bibitem{Borsanyi:2015axp}
S.~Borsanyi, {\em PoS} {\bf LATTICE2015}  (2016),
  \href{http://arxiv.org/abs/1511.06541}{{\ttfamily arXiv:1511.06541}}.

\bibitem{Sexty:2014dxa}
D.~Sexty, {\em PoS} {\bf LATTICE2014},   016  (2014),
  \href{http://arxiv.org/abs/1410.8813}{{\ttfamily arXiv:1410.8813}}.

\bibitem{Gattringer:2014nxa}
C.~Gattringer, {\em PoS} {\bf LATTICE2013},   002  (2014),
  \href{http://arxiv.org/abs/1401.7788}{{\ttfamily arXiv:1401.7788}}.

\bibitem{Aarts:2013lcm}
G.~Aarts, {\em PoS} {\bf LATTICE2012},   017  (2012),
  \href{http://arxiv.org/abs/1302.3028}{{\ttfamily arXiv:1302.3028}}.

\bibitem{Wolff:2010zu}
U.~Wolff, {\em PoS} {\bf LATTICE2010},   020  (2010),
  \href{http://arxiv.org/abs/1009.0657}{{\ttfamily arXiv:1009.0657}}.

\bibitem{deForcrand:2010ys}
P.~de~Forcrand, {\em PoS} {\bf LAT2009},   010  (2009),
  \href{http://arxiv.org/abs/1005.0539}{{\ttfamily arXiv:1005.0539}}.

\bibitem{Chandrasekharan:2008gp}
S.~Chandrasekharan, {\em PoS} {\bf LATTICE2008},   003  (2008),
  \href{http://arxiv.org/abs/0810.2419}{{\ttfamily arXiv:0810.2419}}.

\bibitem{Schubert:2001he}
C.~Schubert, {\em Phys. Rept.} {\bf 355}, 73  (2001).

\bibitem{Gies:2001zp}
H.~Gies and K.~Langfeld, {\em Nucl. Phys.} {\bf B613}, 353  (2001).

\bibitem{Langfeld:2002vy}
K.~Langfeld, L.~Moyaerts and H.~Gies, {\em Nucl. Phys.} {\bf B646}, 158
  (2002).

\bibitem{Gies:2003cv}
H.~Gies, K.~Langfeld and L.~Moyaerts, {\em JHEP} {\bf 06},   018  (2003).

\bibitem{Troyer:2004ge}
M.~Troyer and U.-J. Wiese, {\em Phys. Rev. Lett.} {\bf 94},   170201  (2005).

\bibitem{Gattringer:2012df}
C.~Gattringer and T.~Kloiber, {\em Nucl. Phys.} {\bf B869}, 56  (2013).

\bibitem{Gattringer:2012ap}
C.~Gattringer and T.~Kloiber, {\em Phys. Lett.} {\bf B720}, 210  (2013).

\bibitem{Korzec:2011gh}
T.~Korzec, I.~Vierhaus and U.~Wolff, {\em Comp. Phys. Com.} {\bf 182}, 1477
  (2011).

\bibitem{Rindlisbacher:2016zht}
T.~Rindlisbacher and P.~de~Forcrand  (2016),
  \href{http://arxiv.org/abs/1602.09017}{{\ttfamily arXiv:1602.09017}}.

\bibitem{Prokof'ev:2001zz}
N.~Prokof'ev and B.~Svistunov, {\em Phys.Rev.Lett.} {\bf 87},   160601  (2001).

\bibitem{Wolff:2008km}
U.~Wolff, {\em Nucl. Phys.} {\bf B810}, 491  (2009).

\bibitem{Mercado:2012ue}
Y.~D. Mercado and C.~Gattringer, {\em Nucl.Phys.} {\bf B862}, 737  (2012).

\bibitem{Mercado:2012yf}
Y.~D. Mercado, H.~G. Evertz and C.~Gattringer, {\em Comp. Phys. Com.} {\bf
  183}, 1920  (2012).

\bibitem{Banerjee:2010kc}
D.~Banerjee and S.~Chandrasekharan, {\em Phys. Rev.} {\bf D81},   125007
  (2010).

\bibitem{Wolff:2009ke}
U.~Wolff, {\em Phys. Rev.} {\bf D79},   105002  (2009).

\bibitem{Weisz:2010xx}
P.~Weisz and U.~Wolff, {\em Nucl. Phys.} {\bf B846}, 316  (2011).

\bibitem{Hogervorst:2011zw}
M.~Hogervorst and U.~Wolff, {\em Nucl. Phys.} {\bf B855}, 885  (2012).

\bibitem{Siefert:2014ela}
J.~Siefert and U.~Wolff, {\em Phys. Lett.} {\bf B733}, 11  (2014).

\bibitem{Endres:2006zh}
M.~G. Endres, {\em PoS} {\bf LAT2006},   133  (2006),
  \href{http://arxiv.org/abs/hep-lat/0609037}{{\ttfamily
  arXiv:hep-lat/0609037}}.

\bibitem{Endres:2006xu}
M.~G. Endres, {\em Phys.Rev.} {\bf D75},   065012  (2007).

\bibitem{Wolff:2009kp}
U.~Wolff, {\em Nucl. Phys.} {\bf B824}, 254  (2010), [Erratum: Nucl.
  Phys.834,395(2010)].

\bibitem{Bruckmann:2015sua}
F.~Bruckmann, C.~Gattringer, T.~Kloiber and T.~Sulejmanpasic, {\em Phys. Lett.}
  {\bf B749}, 495  (2015), [Erratum: Phys. Lett.B751,595(2015)].

\bibitem{Bruckmann:2015hua}
F.~Bruckmann, C.~Gattringer, T.~Kloiber and T.~Sulejmanpasic, {\em Phys. Rev.
  Lett.} {\bf 115},   231601  (2015).

\bibitem{Bruckmann:2015uhd}
F.~Bruckmann, C.~Gattringer, T.~Kloiber and T.~Sulejmanpasic, {\em PoS} {\bf
  LATTICE2015}  (2016), \href{http://arxiv.org/abs/1512.05482}{{\ttfamily
  arXiv:1512.05482}}.

\bibitem{Wolff:2010qz}
U.~Wolff, {\em Nucl. Phys.} {\bf B832}, 520  (2010).

\bibitem{Rindlisbacher:2015xku}
T.~Rindlisbacher and P.~Forcrand, {\em PoS} {\bf LATTICE2014}  (2015),
  \href{http://arxiv.org/abs/1512.05684}{{\ttfamily arXiv:1512.05684}}.

\bibitem{Patel:1983sc}
A.~Patel, {\em Nucl.Phys.} {\bf B243},   411  (1984).

\bibitem{DeGrand:1983fk}
T.~A. DeGrand and C.~E. DeTar, {\em Nucl.Phys.} {\bf B225},   590  (1983).

\bibitem{Karsch:1985cb}
F.~Karsch and H.~Wyld, {\em Phys.Rev.Lett.} {\bf 55},   2242  (1985).

\bibitem{Bergner:2013qaa}
G.~Bergner, J.~Langelage and O.~Philipsen, {\em JHEP} {\bf 03},   039  (2014).

\bibitem{Langelage:2014vpa}
J.~Langelage, M.~Neuman and O.~Philipsen, {\em JHEP} {\bf 09},   131  (2014).

\bibitem{Glesaaen:2015vtp}
J.~Glesaaen, M.~Neuman and O.~Philipsen  (2015),
  \href{http://arxiv.org/abs/1512.05195}{{\ttfamily arXiv:1512.05195}}.

\bibitem{Bergner:2015rza}
G.~Bergner, J.~Langelage and O.~Philipsen, {\em JHEP} {\bf 11},   010  (2015).

\bibitem{Mercado:2011ua}
Y.~D. Mercado, H.~G. Evertz and C.~Gattringer, {\em Phys.Rev.Lett.} {\bf 106},
   222001  (2011).

\bibitem{Gattringer:2011gq}
C.~Gattringer, {\em Nucl. Phys.} {\bf B850}, 242  (2011).

\bibitem{Mercado:2013ola}
Y.~D. Mercado, C.~Gattringer and A.~Schmidt, {\em Phys. Rev. Lett.} {\bf 111},
   141601  (2013).

\bibitem{Panero:2005iu}
M.~Panero, {\em JHEP} {\bf 05},   066  (2005).

\bibitem{Korzec:2010sh}
T.~Korzec and U.~Wolff, {\em PoS} {\bf LATTICE2010},   029  (2010),
  \href{http://arxiv.org/abs/1011.1359}{{\ttfamily arXiv:1011.1359}}.

\bibitem{Korzec:2012fa}
T.~Korzec and U.~Wolff, {\em Nucl. Phys.} {\bf B871}, 145  (2013).

\bibitem{Caselle:2014eka}
M.~Caselle, M.~Panero, R.~Pellegrini and D.~Vadacchino, {\em JHEP} {\bf 01},
  105  (2015).

\bibitem{Caselle:2016mqu}
M.~Caselle, M.~Panero and D.~Vadacchino, {\em JHEP} {\bf 02},   180  (2016).

\bibitem{Gattringer:2015nea}
C.~Gattringer, T.~Kloiber and V.~Sazonov, {\em Nucl. Phys.} {\bf B897}, 732
  (2015).

\bibitem{Gattringer:2015baa}
C.~Gattringer, T.~Kloiber and M.~M{\"u}ller-Preussker, {\em Phys. Rev.} {\bf
  D92},   114508  (2015).

\bibitem{Kloiber:2014dfa}
T.~Kloiber and C.~Gattringer, {\em PoS} {\bf LATTICE2014},   345  (2014).

\bibitem{Schmidt:2015cva}
A.~Schmidt, P.~de~Forcrand and C.~Gattringer, {\em PoS} {\bf LATTICE2014},
  209  (2015).

\bibitem{Gattringer:2012jt}
C.~Gattringer and A.~Schmidt, {\em Phys. Rev.} {\bf D86},   094506  (2012).

\bibitem{Mercado:2013yta}
Y.~D. Mercado, C.~Gattringer and A.~Schmidt, {\em Comp. Phys. Com.} {\bf 184},
  1535  (2013).

\bibitem{Rossi:1984cv}
P.~Rossi and U.~Wolff, {\em Nucl. Phys.} {\bf B248},   105  (1984).

\bibitem{Karsch:1988zx}
F.~Karsch and K.~H. M{\"u}tter, {\em Nucl. Phys.} {\bf B313},   541  (1989).

\bibitem{HariDass:2000ca}
N.~D. Hari~Dass, {\em Nucl. Phys. Proc. Suppl.} {\bf 94}, 665  (2001).

\bibitem{HariDass:2000tp}
N.~D. Hari~Dass and D.-S. Shin, {\em Nucl. Phys. Proc. Suppl.} {\bf 94}, 670
  (2001).

\bibitem{HariDass:1999kx}
N.~D. Hari~Dass, {\em Nucl. Phys. Proc. Suppl.} {\bf 83}, 950  (2000).

\bibitem{Cherrington:2007ax}
J.~W. Cherrington, D.~Christensen and I.~Khavkine, {\em Phys. Rev.} {\bf D76},
   094503  (2007).

\bibitem{Cherrington:2007is}
J.~W. Cherrington, {\em Nucl. Phys.} {\bf B794}, 195  (2008).

\bibitem{Cherrington:2008ey}
J.~W. Cherrington and J.~D. Christensen, {\em Nucl. Phys.} {\bf B813}, 370
  (2009).

\bibitem{Cherrington:2009ak}
J.~W. Cherrington, {\em Nucl. Phys.} {\bf B835}, 29  (2010).

\bibitem{Cherrington:2009am}
J.~W. Cherrington, {\em Nucl. Phys.} {\bf B835}, 51  (2010).

\bibitem{deForcrand:2009dh}
P.~de~Forcrand and M.~Fromm, {\em Phys. Rev. Lett.} {\bf 104},   112005
  (2010).

\bibitem{Unger:2011it}
W.~Unger and P.~de~Forcrand, {\em J. Phys.} {\bf G38},   124190  (2011).

\bibitem{deForcrand:2014tha}
P.~de~Forcrand~{\it et al}, {\em Phys. Rev. Lett.} {\bf 113},   152002  (2014).

\bibitem{Salmhofer:1991cc}
M.~Salmhofer, {\em Nucl. Phys.} {\bf B362}, 641  (1991).

\bibitem{Cecile:2007dv}
D.~J. Cecile and S.~Chandrasekharan, {\em Phys. Rev.} {\bf D77},   014506
  (2008).

\bibitem{Chandrasekharan:2002ex}
S.~Chandrasekharan, {\em Phys. Lett.} {\bf B536}, 72  (2002).

\bibitem{DePietri:2007ak}
R.~De~Pietri, A.~Feo, E.~Seiler and I.-O. Stamatescu, {\em Phys. Rev.} {\bf
  D76},   114501  (2007).

\bibitem{Rindlisbacher:2015pea}
T.~Rindlisbacher and P.~de~Forcrand, {\em JHEP} {\bf 02},   051  (2016).

\bibitem{Stamatescu:1980br}
I.~O. Stamatescu, {\em Phys. Rev.} {\bf D25},   1130  (1982).

\bibitem{Gattringer:2015cxh}
C.~Gattringer and V.~Sazonov, {\em Phys. Rev.} {\bf D93},   034505  (2016).

\bibitem{Gattringer:2015pea}
M.~Kniely and C.~Gattringer, {\em PoS} {\bf LATTICE2014},   206  (2015),
  \href{http://arxiv.org/abs/1502.00788}{{\ttfamily arXiv:1502.00788}}.

\bibitem{Gattringer:1999hr}
C.~Gattringer, {\em Nucl. Phys.} {\bf B559}, 539  (1999).

\bibitem{Gattringer:1998cd}
C.~Gattringer, {\em Nucl. Phys.} {\bf B543}, 533  (1999).

\bibitem{Gattringer:1998ri}
C.~Gattringer, {\em Int. J. Mod. Phys.} {\bf A14},   4853  (1999).

\bibitem{Wolff:2008xa}
U.~Wolff, {\em Nucl. Phys.} {\bf B814}, 549  (2009).

\bibitem{Cecile:2008nb}
D.~J. Cecile and S.~Chandrasekharan, {\em Phys. Rev.} {\bf D77},   054502
  (2008).

\bibitem{Endres:2012vd}
M.~G. Endres, {\em Phys. Rev.} {\bf A85},   063624  (2012).

\bibitem{Chandrasekharan:2009wc}
S.~Chandrasekharan, {\em Phys. Rev.} {\bf D82},   025007  (2010).

\bibitem{Chandrasekharan:2011mn}
S.~Chandrasekharan and A.~Li, {\em Phys. Rev. Lett.} {\bf 108},   140404
  (2012).

\bibitem{Chandrasekharan:2012va}
S.~Chandrasekharan and A.~Li, {\em Phys. Rev.} {\bf D85},   091502  (2012).

\bibitem{Chandrasekharan:2012fk}
S.~Chandrasekharan, {\em Phys. Rev.} {\bf D86},   021701  (2012).

\bibitem{Wang:2001ab}
F.~Wang and D.~P. Landau, {\em Phys. Rev. Lett.} {\bf 86}, 2050  (2001).

\bibitem{Xu:2007aa}
J.~Xu and H.-R. Ma, {\em Phys. Rev. E} {\bf 75},   041115  (2007).

\bibitem{Sinha:2007aa}
S.~Sinha and S.~Kumar~Roy, {\em Phys.Lett.} {\bf A373}, 308  (2009).

\bibitem{Berg:1992qua}
B.~Berg and T.~Neuhaus, {\em Phys.Rev.Lett.} {\bf 68}, 9  (1992).

\bibitem{Langfeld:2012ah}
K.~Langfeld, B.~Lucini and A.~Rago, {\em Phys.Rev.Lett.} {\bf 109},   111601
  (2012).

\bibitem{Langfeld:2015fua}
K.~Langfeld, B.~Lucini, R.~Pellegrini and A.~Rago  (2015),
  \href{http://arxiv.org/abs/1509.08391}{{\ttfamily arXiv:1509.08391}}.

\bibitem{Gattringer:2015lra}
C.~Gattringer and P.~T{\"o}rek, {\em Phys. Lett.} {\bf B747}, 545  (2015).

\bibitem{robbins1951}
H.~Robbins and S.~Monro, {\em Ann. Math. Statist.} {\bf 22}, 400 (09 1951).

\bibitem{Langfeld:2013xbf}
K.~Langfeld and J.~M. Pawlowski, {\em Phys. Rev.} {\bf D88},   071502  (2013).

\bibitem{Langfeld:2014nta}
K.~Langfeld and B.~Lucini, {\em Phys. Rev.} {\bf D90},   094502  (2014).

\bibitem{Gocksch:1988iz}
A.~Gocksch, {\em Phys.Rev.Lett.} {\bf 61},   2054  (1988).

\bibitem{Azcoiti:2002nq}
V.~Azcoiti~{\it et al}, {\em Nucl.Phys.Proc.Suppl.} {\bf 119}, 1009  (2003).

\bibitem{Azcoiti:2011ei}
V.~Azcoiti, E.~Follana and A.~Vaquero, {\em Nucl.Phys.} {\bf B851}, 420
  (2011).

\bibitem{Mercado:2014dva}
Y.~Delgado~Mercado, P.~T{\"o}rek and C.~Gattringer, {\em PoS} {\bf
  LATTICE2014},   203  (2015).

\bibitem{Gattringer:2015eey}
C.~Gattringer~{\it et al}, {\em POS} {\bf LATTICE2015},   194  (2015),
  \href{http://arxiv.org/abs/1511.07176}{{\ttfamily arXiv:1511.07176}}.

\end{thebibliography}

\end{document}